\documentclass[%
 reprint,
superscriptaddress,
longbibliography,
bibnotes,
 amsmath,amssymb,
 aps,
pre,
floatfix,
]{revtex4-2}

\usepackage{graphicx}% Include figure files
\graphicspath{ {./Figures/} }

\usepackage{dcolumn}% Align table columns on decimal point
\usepackage{bm}% bold math
\usepackage[normalem]{ulem}
\usepackage{xcolor}
\definecolor{OliveGreen}{rgb}{0,0.6,0}
\definecolor{auburn}{rgb}{0.43, 0.21, 0.1}
\definecolor{BlueViolet}{rgb}{0.54, 0.17, 0.89}
\definecolor{HokieOrange}{RGB}{232, 119, 34}
\definecolor{HokieMaroon}{RGB}{134, 31, 65}
\definecolor{Black}{RGB}{0, 0, 0}
\definecolor{HokieMaroon}{RGB}{134, 31, 65}

\usepackage{appendix}
\usepackage{chngcntr}
\usepackage{etoolbox}
\usepackage{lipsum}
\AtBeginEnvironment{appendices}{%
\addcontentsline{toc}{section}{Appendices}
\counterwithin{equation}{section}
\counterwithin{table}{section}
}

\begin{document}

%\preprint{APS/123-QED}

\title{Entropy-Driven Microstructure Evolution Predicted with the Steepest-Entropy-Ascent Quantum Thermodynamic Framework}% Force line breaks with \\

\author{Jared McDonald}
\altaffiliation{jmcdonald@vt.edu (J. McDonald)}
\affiliation{Materials Science and Engineering Department, Virginia Tech, Blacksburg, VA 24061, USA}
\author{Michael R. von Spakovsky}
\altaffiliation{vonspako@vt.edu (M.R. von Spakovsky)}
\affiliation{Mechanical Engineering Department, Virginia Tech, Blacksburg, VA 24061, USA}
\author{William T. Reynolds Jr.}
\altaffiliation{reynolds@vt.edu (W. T. Reynolds Jr.)}
\affiliation{Materials Science and Engineering Department, Virginia Tech, Blacksburg, VA 24061, USA}

%\date{\today}
\date{2021-08-22}

\begin{abstract}
A Potts model and the Replica Exchange Wang-Landau algorithm is used to construct an energy landscape for a crystalline solid containing surfaces and grain boundaries. The energy landscape is applied to an equation of motion from the steepest-entropy-ascent quantum thermodynamic (SEAQT) framework to explore the kinetics of three distinct kinds of microstructural evolution: polycrystalline sintering, precipitate coarsening, and grain growth. The steepest entropy ascent postulate predicts unique kinetic paths for these non-equilibrium processes without needing any detailed information about the underlying physical mechanisms of the processes. A method is also proposed for associating the kinetic path in state space to a set of smoothly evolving microstructural descriptors. The SEAQT-predicted kinetics agree well with available experimental kinetics for ZrO$_2$ sintering, Al$_3$Li precipitate coarsening, and grain growth in nanocrystalline Pd. The computational cost associated with calculating the energy landscape needed by the approach is comparable to a Monte Carlo simulation. However, the subsequent kinetic calculations from the SEAQT equation of motion are quite modest and save considerable computational resources by obviating the need for averaging multiple kinetic Monte Carlo runs.
\end{abstract}

%\pacs{Valid PACS appear here}% PACS, the Physics and Astronomy
                             % Classification Scheme.
\maketitle

\section{Introduction}

The impetus for microstructural evolution lies in one of Clausius's seminal statements of the second law of thermodynamics: the entropy of an isolated system at constant energy tends to a maximum. Historically, this maximum entropy principle was rarely used in materials science because it is impossible to relate entropy directly to measurable microstructural parameters. Instead, changes during processes like particle sintering, grain growth, and precipitate coarsening were typically modeled by the conjugate principle of minimizing energy at constant entropy. The kinetics of these processes were expressed as a linear function of a driving force typically taken to be a local free-energy change associated with reducing the area of surfaces and grain boundaries.  

In contrast to such kinetic descriptions, the steepest-entropy-ascent quantum thermodynamic (SEAQT) framework provides a practical vehicle for applying Clausius's maximum entropy principle.  The framework identifies unique kinetic paths to stable equilibrium without the need for the usual near- and local-equilibrium assumptions. The SEAQT approach is based upon entropy calculated directly from a discrete energy landscape that covers all possible microstructures of a system. The energy landscape is determined from an appropriate model that depends upon the nature of the physical system \cite{Li2016a,Li2016b,Li2016c,Li2018,Li2018steepest,Li2018steepest,Li2017study, Li2018multiscale,yamada2018method,Yamada2019,yamada2019kineticpartI,yamada2019spin,yamada2020kineticpartII,jhon2020,cano2015steepest,kusaba2019,vonSpakovsky2020}. The model can either be quantum mechanically-based or quantum mechanically-inspired (e.g., solid-state, Ising, Heisenberg, or Potts models).  

The contribution presented here applies the SEAQT framework to an energy landscape to describe the kinetics of three kinds of microstructural evolution to demonstrate the generality and flexibility of the approach. The energy landscape is based on a Potts model, which is developed using the Replica Exchange Wang-Landau algorithm  \cite{Vogel2013,Vogel2014} with a Hamiltonian defined for a solid with surfaces and grain boundaries. The algorithm is used to numerically generate the energy landscape and corresponding density of states for the system. The state of the system is expressed as a probability density distribution at each instant of time over the energy levels of the system's energy landscape \footnote{Note that the basis for such a probability density distribution in the quantal formulation of the SEAQT framework is the density or so-called ``state'' operator, which is based on a homogeneous ensemble \cite{hatsopoulos1976-III}}, and expected values of the energy and entropy are calculated directly from these time-dependent probabilities. The probability density distributions are uniquely predicted by the SEAQT equation of motion, which provides the time evolution of the probabilities that describe the occupancy of the energy levels of the landscape and, thus, the non-equilibrium state of the system in time.

Different starting points or initial conditions on the energy landscape give rise to qualitatively and quantitatively different kinetic paths along which the system evolves. The results for three different initial conditions are given here to demonstrate the phenomenological behavior corresponding to polycrystalline sintering, precipitate coarsening (or Ostwald ripening), and grain growth. The results include time evolutions of the system microstructure as well as the microstructure's average grain size, the number of grain boundaries and surface boundaries, and the relative density. Even though the three evolutions involve distinctly different phenomenological behaviors, they are obtained with a single model and a single energy landscape without reference to kinetic mechanisms or assumed limiting rates; state evolution in each case is driven simply by the principle of steepest entropy ascent {\color{black}(SEA)}.  \textcolor{Black}{This principle has been postulated as a fundamental law of nature~\cite{Martyushev2021}} and is used by the SEAQT equation of motion~\cite{Beretta2020} to maximize the entropy production at each instant along the non-equilibrium path the system takes through state space

\textcolor{Black}{It is significant that the {\color{black}SEA} principle in this framework is not merely a constrained optimization with an objective function and a set of decision variables, but it is instead the application of a variational principle that leads to a unique thermodynamic path through state space. The former, for example, provides for a maximum growth velocity as in the case of Martyushev’s application of the MEP to dendrite growth~\cite{Martyushev2013} or to a stability point uniquely identified as a minimum/maximum in the entropy production as in Kirkaldy’s application to eutectic spacing~\cite{KirkaldySharma1980} but has nothing to say about the unique non-equilibrium transient thermodynamic path taken. In contrast, the variational principle of the SEAQT framework assumes that nature always seeks a thermodynamic path that satisfies an extremum. This is analogous to the variational principle used in classical mechanics that determines the unique trajectory of a particle from an infinite number of possible trajectories by finding the ``least action'' (i.e., minimizing the difference of the kinetic and potential energies represented by the Lagrangian). The result is the set of Euler-Lagrange equations and as a consequence the equation of motion of Newtonian physics. It is such a variational principle based on {\color{black}SEA} that leads to the SEAQT equation of motion and the ability to predict the unique non-equilibrium transient thermodynamic path taken by a system.}

The question, of course, arises as to why the extremum in this case should be the maximization as opposed to the minimization of the entropy production. At first glance, the two would seem to contradict each other. However, they do not. As has been very clearly shown, one can arrive at both linear and nonlinear non-equilibrium thermodynamics from Ziegler’s maximum entropy production principle~\cite{ziegler1963some,ziegler1957thermodynamik,ziegler1983chemical,ziegler1983introduction,ziegler1987principle}, which yields as a particular case Onsager’s linear result. This contrasts with Prigogine’s minimum entropy production principle~\cite{Prigogine1967}, which is a particular case of the Onsager-Gyarmati principle of linear non-equilibrium thermodynamics in which a stationary process is in the presence of free thermodynamic forces (e.g., a free boundary or indeterminate diffusion as in reference~\cite{KirkaldySharma1980}). However, there is evidence to suggest that in nonstationary processes and in stationary processes well-removed from equilibrium nature chooses to operate with fixed forces at any given instant of time and, thus, maximizes the entropy production at each moment~\cite{yamada2020kineticpartII,Martyushev2021}. It is this idea which forms the basis of Ziegler’s principle as well as that of Beretta~\cite{beretta2005generalPhD,Beretta1984, Beretta1985, beretta2006nonlinear, beretta2009nonlinear, beretta2014steepest}: that the direction nature chooses at every instant of time is that of steepest entropy ascent.

{\color{black}A final comment about the generality of the SEA principle is illustrated with the work of Kirkaldy \cite{Kirkaldy1960,Kirkaldy1959,Kirkaldy1964} in which, using a variational principle, he applies the minimum entropy principle  to diffusional growth in solid-solid transformations. Cahn and Mullins \cite{CahnMullins1964} challenge the generality of this principle and Kirkaldy's use of it and do so by using two simple examples: i) one-dimensional steady state heat conduction and ii) one-dimensional steady state mass diffusion in the presence of an externally maintained temperature gradient. In both examples, Cahn and Mullins correctly show that in this particular case, minimum entropy production does not correspond to the correct temperature or concentration profile that would result from the transient thermal and mass diffusion equations and Fourier's and Fick's laws. Thus, for the case of thermal diffusion, the temperature profile, which should be linear at steady state, is shown by Cahn and Mullins, using a variational principle, to be logarithmic for the case of minimum entropy production. However, all that this particular example and the other one demonstrate is that minimum entropy production as implemented by Cahn and Mullins via a variational principle does not correspond to the correct steady state profile, which can, however, be determined via the SEA principle as implemented in the SEAQT equation of motion. In fact, this is done for the first of these two examples by Li, von Spakovsky, and Hin in Appendix B of \cite{Li2018steepest} where using the SEAQT equation of motion, the authors predict the correct steady state linear temperature profile for the case of constant thermal conductivity. They as well show that when the thermal conductivity is not constant, the steady state temperature profile is somewhat nonlinear as would be expected. This is done using the SEA principle $\textit{only}$. No assumption of a particular kinetic mechanism, i.e., Fourier's law as used in the transient thermal diffusion equation, is made. For the case of mass diffusion in the presence of a fixed temperature gradient, the expected steady state linear  concentration profile is predicted by Li and von Spakovsky in \cite{Li2016b} using the SEAQT equation of motion. Again, this is done without an $\textit{a priori}$ assumption of a particular kinetic mechanism, i.e., in this case, Fick's law as employed in the transient mass diffusion equation.
	
Finally, the} remainder of the paper is organized as follows. Section \ref{REWL_sec:level2} describes the energy landscape and the Replica Exchange Wang-Landau method for developing this landscape. Section \ref{SEAQT_sec:level2} presents the SEAQT equation of motion and discusses how it is formulated for this application, and Section \ref{State2Microstructure_sec:level3} outlines how the system's state space is linked to the system's microstructure. Sections \ref{Results_subsec:Sintering}, \ref{Results_subsec:Ostwald}, and \ref{Results_subsec:GG} present the results of the models for sintering, precipitate coarsening, and grain growth, respectively, and Section \ref{conclusions_sec:level1} provides some summary conclusions.

%Use of the SEAQT equation of motion will allow for the derivation of the unique kinetic path of the agglomerating system derived solely from on the inputted initial conditions \cite{Beretta1985,Beretta2009,Beretta2006}. 

\section{Method \label{method_sec:level1}}

\subsection{{Energy Landscape} \label{REWL_sec:level2}}

The system microstructure is described by a 2-dimensional grid of pixels \cite{Zhang2019,Hara2015,Bjork2014,Tikare2010,Braginsky2005} whose energy is given by a $q$-spin Potts model.  This model defines a variable, $q$, that monitors the spin of each pixel in the system. The $q$ integers can range from 0 to several hundred, depending upon what physical entity the $q$ phases represent. In this work, each location has an associated integer $q$ value which represents either a void (when $q=0$) or grain orientation ($q \geqslant 1$). The larger the maximum number of $q$ values is, the larger the number of allowed grain orientations. Surface energy arises when a void pixel is adjacent to a solid pixel (i.e., when a pixel with $q=0$ is adjacent to a pixel with $q \geqslant 1$). Grain boundary energy arises between pixels with different positive $q$ values. 

The system is represented by a square $L \times L$ lattice where $L$ is the linear size in pixels. The total energy, $E$, of the system is the sum of the energies of all the surface and grain boundaries and is represented mathematically by the following Potts model interaction Hamiltonian
\cite{Zhang2019,Hara2015,Bjork2014,Tikare2010,Braginsky2005}: % Jared: I changed the symbols in the following equations to make them consistent with descriptions in the text and to avoid using some symbols multiple times for different purposes. please check equations (1), (3)-(9) to make sure they are still correct

%VERSION WITH OLD VARIABLES 
%\begin{equation} \label{TotalEnergy}
%E = \frac {1} {2}\underset {i = 1} {\overset {N} {\sum}}\underset {j = 1} {\overset {n} {\sum}} \; J \left(1 - \delta (q_i, q_j) \right) 
%\end{equation}
%VERSION WITH OLD VARIABLES 
%{\color{black}In this equation, $E$ is determined} by summing over the number of sites (pixels), $N\; (= L^2)$, and the number of neighbors {\color{black}to each site, $n$}. The Potts coupling constant or interaction {\color{black}energy}, $J$, switches between a surface energy, $\gamma_{s}$, and a grain boundary energy, $\gamma_{gb}$, depending upon the identities of $q_i$ and $q_i$. The $\delta$ on the right side represents the Kronecker delta which returns a value of $1$ if the neighboring $j$th grain to $i$ is of the same orientation {\color{black}(i.e., $q_i = q_j$) or} $0$ if it is not ($q_i \neq q_j$) \cite{Zhang2019,Hara2015,Bjork2014,Tikare2010,Braginsky2005}.

\begin{equation} 
E = \frac{1}{2}\, \underset {n = 1} {\overset {N} {\sum}} \, \underset {z = 1} {\overset {Z} {\sum}} \; J \left(1 - \delta (q_n, q_z) \right)
\label{TotalEnergy}
\end{equation}
In this equation, $E$ is determined by summing over the number of lattice sites (pixels), $N\; (= L^2)$, and the number of neighbors to each site, $Z$. The Potts coupling constant or interaction energy, $J$, switches between a surface energy, $\gamma_{s}$, and a grain boundary energy, $\gamma_{gb}$, depending upon the identities of $q_n$ and $q_z$. The $\delta$ on the right side represents the Kronecker delta which returns a value of $1$ if the neighboring $z^{th}$ grain to $n$ is of the same orientation (i.e., $q_n = q_z$), or $0$ if it is not ($q_n \neq q_z$) \cite{Zhang2019,Hara2015,Bjork2014,Tikare2010,Braginsky2005}. Thus, the only contributions to the total energy come from boundaries in the system.

Equation (\ref{TotalEnergy}) gives the energy of any arbitrary configuration or state of solid grains bounded by surfaces and/or grain boundaries.  The energy landscape, or energy eigenstructure, represents all the energies of all possible system configurations.  For lattices of appreciable size, many of these configurations have the same energy; stated conversely, most energy levels are degenerate. In order to calculate system properties like the entropy, it is essential to know the degeneracy of each energy level. This information is typically represented as a density of states for each level.

The density of states for the system can be obtained from the Wang-Landau method \cite{WangLandau2001a,WangLandau2001b}. The Wang-Landau method is a non-Marcovian approach for estimating the degeneracies from a flat histogram generated via a Monte Carlo walk through all the possible energy levels of the system. The algorithm estimates the degeneracies from the fact that Monte Carlo transitions between individual energy levels occur with a probability given by $1/g(E_j)$, where $g(E_j)$ is the estimated degeneracy of the $E_j$ energy level. Repeated Monte Carlo sweeps through the energy spectrum refines the accuracy of the estimates. The ``replica exchange'' \cite{Vogel2013,Vogel2014} variant of the Wang-Landau method greatly accelerates the algorithm by subdividing the energy spectrum into multiple windows, utilizing multiple Monte Carlo walkers over the energy windows, and passing information among them. The basis of the Replica Exchange Wang-Landau code employed here is given in Vogel, Li, and Landau \cite{Vogel2018}.

\begin{figure*}[h!]
\begin{center}
\includegraphics[width=1\textwidth]{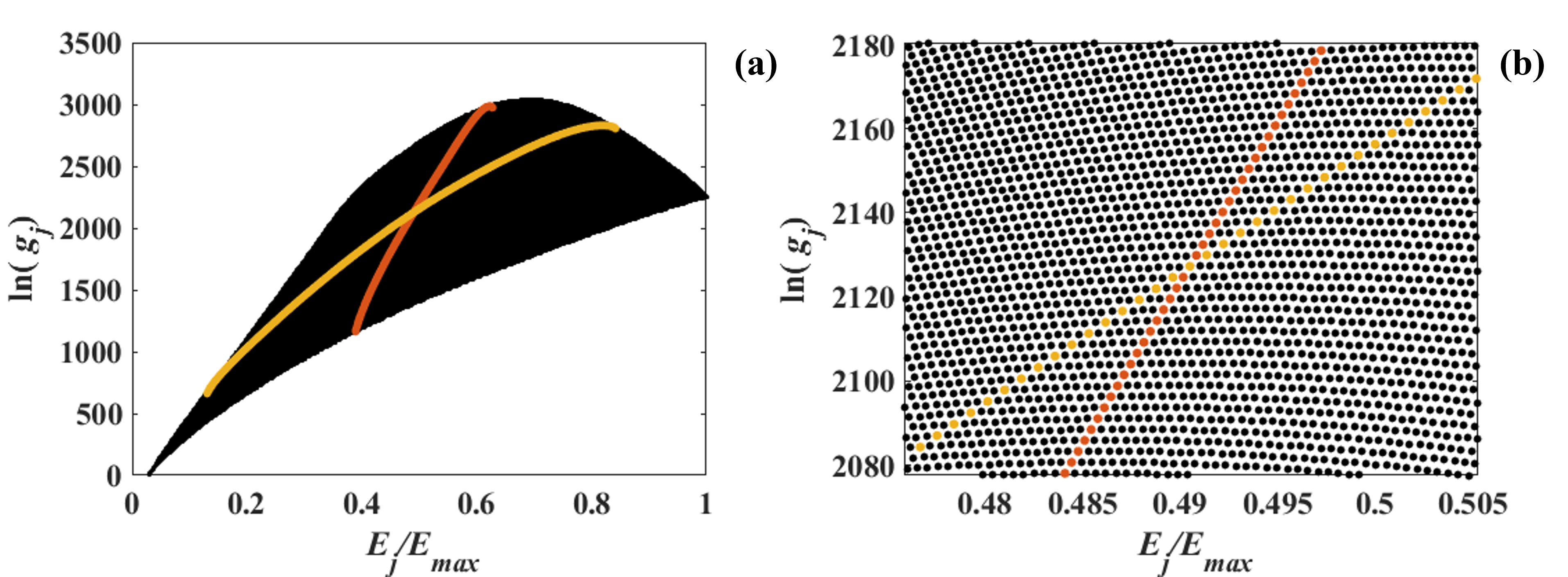}
\caption{\textcolor{Black}{Density of states calculated with the Replica Exchange, Wang-Landau algorithm for a $34 \times 34$ lattice consisting of 50\% solid with surfaces and grain boundaries. There are 629,997 discrete energy levels for the system. The horizontal axis is scaled by the energy of the maximum level, and the vertical axis is the natural log of the degeneracy of the $E_j$ energy eigenlevels. (a) shows the entire energy landscape; the individual energy eigenlevels are represented by points that are too close to differentiate so that the density of states appears like a solid, continuous region. (b) is a greatly enlarged segment of (a) that reveals the individual energy levels. The energy eigenlevels are arranged in arcs that correspond to iso-grain-boundary-areas (an example arc of iso-grain-boundary-area is highlighted in orange) and iso-surface-areas (an example arc of iso-surface-area} is highlighted in yellow).}
\label{fig:DOS}
\end{center}
\end{figure*}

The density of states calculated with the Replica Exchange Wang-Landau algorithm for an $L \times L$ lattice with $L=34$ is shown in Figure \ref{fig:DOS}. \textcolor{Black}{The lattice was chosen to have 50\% of the pixels as voids ($q=0$) and the remaining 50\% with values of} $q \geqslant 1$. The plot represents the natural log of the number of states or configurations as a function of the state energy. For this energy landscape, the surface energy and the grain boundary energy are assumed to be isotropic, and there are a maximum of 50 distinct grain orientations.  This figure represents all the energies of all the possible states (configurations) of the physical system and the degeneracy of each energy level. \textcolor{Black}{Each eigenenergy along the abscissa has a unique degeneracy.}

%\begin{figure*}[h!]
%\begin{center}
%\includegraphics[width=0.6\textwidth]{PSD.pdf}
%\caption{Fictitious caption to some random figure.}
%\label{fig:PSD}
%\end{center}
%\end{figure*}

\subsection{{SEAQT Equation of Motion} \label{SEAQT_sec:level2}}

The SEAQT equation of motion is used to predict the non-equilibrium thermodynamic behavior of the system. This equation requires no {\em a priori} knowledge of the kinetic mechanisms involved because it predicts the evolution of system properties on the principle of steepest entropy ascent. \textcolor{Black}{The energy transitions between levels of the landscape provide the underlying first-principle-basis for the kinetic phenomena that a given system experiences. It is these transitions that the SEAQT equation of motion captures with the principle of steepest entropy ascent (or equivalently, maximum entropy production), satisfying in the process the first and second laws of thermodynamics and the postulates of quantum mechanics provided the quantum mechanical features of the system have been included in the energy landscape. In the case of our application here, the quantum features are not needed and, thus, are not included. From a phenomenological and continuum standpoint, the constitutive laws and kinetic mechanisms typically used by traditional deterministic and stochastic material science models have an underlying second law of thermodynamics basis rooted in Onsager’s linear theory of nonequilibrium thermodynamics in which a direct correlation between paired generalized forces and fluxes provides an estimate of the rate of entropy production. That linear phenomenological continuum approach cannot, however, from purely thermodynamic considerations, be shown to be applicable in regions other than those close to equilibrium. That limitation does not apply to the SEAQT framework, which has been shown thermodynamically to be applicable throughout the non-equilibrium region~\cite{Beretta1984, Li2016b, Li2018}. Another advantage is  the SEAQT framework inherently captures the effects of coupled~\cite{Li2018steepest, yamada2019spin} and concurrent~\cite{yamada2020kineticpartII} phenomena when constitutive relationships such as Fick’s law, Fourier’s law, Ohm’s law, \em etc., and simple rate-limiting models break down.  The approach calculates the entropy production directly from the distribution of the system energy among the levels of a discrete energy landscape so there is no need to build a microstructural model with field equations involving fluxes and the corresponding thermodynamic forces associated with gradients in chemical potential, temperature, or electric potential. As a consequence, the accuracy of the SEAQT approach comes entirely from the details of the energy landscape rather than the details of assumed kinetic mechanisms, which may or may not be generally applicable. The SEAQT framework uses one universal kinetic model of energy transitions captured by an equation of motion that satisfies the laws and postulates of thermodynamics and quantum mechanics. Changing the application, i.e., system, simply requires building a different energy landscape.}

For the case of a simple quantum system, the equation of motion is expressed as \cite{Beretta2006,Beretta2009,Li2016d}
 
\begin{equation} 
\color{black}
\frac {d \hat{\rho}} {dt}=\frac {1} {i \hbar}[\hat{\rho},\hat{H}]+\frac {1} {\tau(\hat{\rho})} {\hat D(\hat{\rho})} \label{EOM1}
\end{equation}
In this expression, $t$ is time, $\hbar$ the modified Planck constant, $\hat{H}$ the Hamiltonian operator, $\hat{D}$ the dissipation operator, $\tau$ the relaxation parameter, and $\hat{\rho}$ the density operator, which represents the thermodynamic state of the system (i.e., the distribution of eigenstates that comprise the thermodynamic state) at each instant of time. Note that $\left[\cdot\right]$ represents the Poisson bracket. The term on the left-hand side of the equation and the first term on the right side, the so-called symplectic term, constitute the time-dependent part of the von Neumann form of the Schr\"odinger equation of motion used in quantum mechanics to predict the reversible evolution of pure states (i.e., zero-entropy states). The second term on the right is there to capture evolutions involving the non-zero-entropy states of irreversible processes. 

%\pagecolor{black}

Now, since the energy landscape considered here is only quantum-inspired and as a result contains no quantum information, the density operator reduces to a probability density distribution and the symplectic term is zero (because there are no quantum correlations) and, thus, $\hat{H}$ and $\hat{\rho}$, which is diagonal in the energy eigenvalue basis of the Hamiltonian, commute  \cite{Li2016a,Li2016b,Li2018,Beretta2006,Beretta2009,Li2016d}. Furthermore, $\hat{D}$, which was originally postulated by Beretta \cite{Beretta1984,Beretta1985}, can be derived via a variational principle that preserves the energy and occupational probabilities by a constrained-gradient descent in Hilbert space along the direction of steepest entropy ascent at each instant of time. For the case considered here when the only two generators of the motion are the Hamiltonian and the identity operators, the equation of motion (Eq. (\ref{EOM1}) reduces to \cite{Beretta2006,Beretta2009,Li2016d}:

%VERSION WITH OLD VARIABLES
%\begin{equation} 
%\frac{dp_j}{dt}=\frac {1} {\tau}\frac{\left| 
%\begin{array}{ccc}
% -p_j \text{ln}\frac{p_j}{g_j} & p_j & {\epsilon}_j p_j \\
% \langle s \rangle & 1 & \langle e \rangle \\
% \langle es \rangle & \langle e \rangle & \langle e^2 \rangle \\
%\end{array}
%\right|}{\left|
%\begin{array}{cc}
% 1 & \langle e \rangle \\
%  \langle e \rangle & \langle e^2 \rangle \\
%\end{array}
%\right|}
%\end{equation} 

\begin{equation} 
\color{black}
\frac{dp_j}{dt}=\frac {1} {\tau}\frac{\left| 
\begin{array}{ccc}
 -p_j \ln \frac{p_j}{g_j} & p_j & {E}_j\, p_j \\
 \langle S \rangle & 1 & \langle E \rangle \\
 \langle E\,S \rangle & \langle E \rangle & \langle E^2 \rangle \\
\end{array}
\right|}{\left|
\begin{array}{cc}
 1 & \langle E \rangle \\
  \langle E \rangle & \langle E^2 \rangle \\
\end{array}
\right|}
\end{equation} 

Here, $p_j$ represents the occupation probability of the $j^{th}$ energy eigenlevel, $E_j$, and $g_j$ its associated degeneracy. The degeneracy is the number of possible system configurations for a given energy eigenvalue. In the SEAQT framework, the von Neumann form for entropy, $S_j = -p_j \text{ln}\frac{p_j}{g_j}$, is used because it satisfies the necessary characteristics for the entropy required by thermodynamics \cite{Gyftopoulos1997} and it provides a simple means of directly calculating the entropy of the system in any of its possible states. Additionally, $\langle \cdot \rangle$ represents the expectation value of a property of the system such as the energy, $E$, the entropy, $S$, the energy squared, $E^2$, or the product of the energy and entropy \cite{Beretta2006,Beretta2009,Li2016d}, e.g., 
%VERSION WITH OLD VARIABLES
%\begin {equation}
%\langle e^2 \rangle = \underset {i} {{\sum}}\phantom{l}{\epsilon}_i^2 p_i 
%\end {equation}
% 
%\begin {equation}
%\langle es \rangle = \underset {i} {{\sum}}\phantom{l}{\epsilon}_i p_i \text{ln}\frac{p_i}{g_i}
%\end {equation}
\begin {equation}
\color{black}
\langle E^2 \rangle = \underset {j} {{\sum}}\phantom{l}{E}_j^2\, p_j 
\end {equation}
 
\begin {equation}
\color{black}
\langle E\, S \rangle = \underset {j} {{\sum}}\phantom{l}{E}_j \, p_j \text{ln}\frac{p_j}{g_j}
\end {equation}

Although the equation of motion is only applicable to an isolated system, any set of systems may be treated as an isolated composite of subsystems. This enables interactions --- such as the exchange of heat and mass --- among subsystems to be taken into account. For example, the equation of motion for an isolated composite of two subsystems $A$ and $B$ experiencing a heat interaction is given by \cite{Li2016a}:
 
%VERSION WITH OLD VARIABLES
%\begin{equation} 
%\frac{dp_j}{dt}=\frac {1} {\tau}\frac{\left| 
%\begin{array}{cccc}
% -p_j^A \text{ln}\frac{p_j^A}{g_j^A} & p_j^A &0 & {\epsilon}_j^A p_j^A \\
% \langle s \rangle^A & 1 & 0 &\langle e \rangle^A \\
% \langle s \rangle^B & 0 & 1 &\langle e \rangle^B \\
% \langle es \rangle & \langle e \rangle^A & \langle e \rangle^B & \langle e^2 \rangle \\
%\end{array}
%\right|}{\left|
%\begin{array}{ccc}
% 1 & 0& 
%\langle e \rangle^A \\
%  0&1&\langle e \rangle^B \\
%   \langle e \rangle^A &\langle e \rangle^B &\langle e^2 \rangle\\
%\end{array}
%\right|}
%\end{equation}
\begin{equation} 
\color{black}
\frac{dp_j}{dt}=\frac {1} {\tau}\frac{\left| 
\begin{array}{cccc}
 -p_j^A \text{ln}\frac{p_j^A}{g_j^A} & p_j^A &0 & {E}_j^A p_j^A \\
 \langle S^A \rangle & 1 & 0 &\langle E^A \rangle \\
 \langle S^B \rangle & 0 & 1 &\langle E^B \rangle \\
 \langle E\,S \rangle & \langle E^A \rangle & \langle E^B \rangle & \langle E^2 \rangle \\
\end{array}
\right|}{\left|
\begin{array}{ccc}
 1 & 0& 
\langle E^A \rangle \\
  0&1&\langle E^B \rangle \\
   \langle E^A \rangle &\langle E^B \rangle &\langle E^2 \rangle\\
\end{array}
\right|}
\end{equation}

Using the cofactors of the first line of the numerator's determinant, $C_1$, $C_2^A$, and $C_3$, and assuming the hypoequilibrium condition developed in \cite{Li2016a}, the equation of motion for subsystem $A$ can be written compactly as \cite{Li2016a}:
 
%VERSION WITH OLD VARIABLES
%\begin{align}
%\frac {dp_j^A}{dt} &= p_j^A \left(-\text{ln}\frac{p_j^A}{g_j^A}-\frac{C_2^A}{C_1}-{\epsilon}_j^A\frac{C_3}{C_1}\right) \nonumber \\
%& = p_j\left[(s_j^A - \langle s \rangle^A)-({\epsilon}_j^A - \langle {\epsilon}\rangle^A)\frac{C_3}{C_1}\right] \\
%& = 
%p_j\left[(s_j^A - \langle s \rangle^A)-({\epsilon}_j^A - \langle {\epsilon}\rangle^A){\beta}\right] \nonumber
%\end{align}
\begin{align}
\frac {dp_j^A}{dt^*} & = p_j^A \left(-\text{ln}\frac{p_j^A}{g_j^A}-\frac{C_2^A}{C_1}-{E}_j^A\frac{C_3}{C_1}\right) \nonumber \\
&   = p_j\left[(S_j^A - \langle S^A \rangle)-({E}_j^A - \langle {E}^A\rangle)\frac{C_3}{C_1}\right] %\\
% &   = 
% p_j\left[(S_j^A - \langle S \rangle^A)-({E}_j^A - \langle {E}\rangle^A){\beta}\right] \nonumber
\end{align}
% fixed the definition for t* (Bill)
where the dimensionless time
 \textcolor{black}{$t^*= \int_0^t \frac{1}{\tau(\vec{p}(t'))}dt'$}
is used to replace $t$ and $\tau$. \textcolor{Black}{The relaxation parameter, $\tau$, from the equation of motion describes the system’s dynamic speed along the kinetic path from the initial state to stable equilibrium. In the most general case, $\tau$ is a function of the time-dependent occupation probabilities $p_j$ represented by the vector $\vec{p}$.  It can be estimated from {\em ab-initio} calculations of the transition rate among energy levels in the system, or in the absence of such detailed information, it can be used as a fitting parameter to scale the predicted SEAQT kinetics to experimental data~\cite{yamada2019kineticpartI}.}
 
If the size of subsystem $B$ is assumed to be significantly larger than $A$, subsystem $B$ can be treated as a reservoir, denoted by $R$, and the previous equation reduces to \cite{Li2016a, Li2018}
 
%VERSION WITH OLD VARIABLES
%\begin{align} \label{eqnEOM}
%\frac {dp_j^A}{dt} = p_j\left[(s_j^A - \langle s \rangle^A)-({\epsilon}_j^A - \langle {\epsilon}\rangle^A){\beta}^R\right]
%\end{align}
\begin{align} \label{eqnEOM}
\frac {dp_j^A}{dt^*} = p_j\left[(S_j^A - \langle S^A \rangle)-({E}_j^A - \langle {E}^A\rangle){\beta}^R\right]
\end{align}
where $\beta^R = C_3/C_1$ reduces to $\frac{1}{k_b T^R}$ with $T^R$ representing the temperature of the reservoir. With this formulation, the constant-temperature kinetic processes can be simulated \cite{Li2016a, Li2016b}. 

\textcolor{Black}{Eq.~(\ref{eqnEOM}) is a system of equations that is solved simultaneously to yield the time-dependent occupation probabilities of the energy eigenlevels. These probabilities collectively describe {\em one} unique path, i.e., the steepest-entropy-ascent path. This is analogous to solving an Euler-Lagrange equation; the solution is guaranteed to yield the extremal path that one is seeking.  This is an important advantage of the SEAQT framework.  Because the path is in state space, it already accounts for all possible configurations – including all the local spatial and temporal fluctuations that appear in microstructural space. Solving Eq.~(\ref{eqnEOM}) results in the globally maximum entropy production and is, thus, guaranteed to end at equilibrium and cannot get trapped “along the way” in microstructures associated with local free-energy minima. The solution provides the non-equilibrium thermodynamic state probability distribution at each instant of time and culminates with the probability distribution at equilibrium. These distributions are then used to calculate the thermodynamic, transport, and material properties of the system at each instant of time. The system of equations, Eq.~(\ref{eqnEOM}), varies in size but can be quite large as is the case for our particular problem in which there are 629,997 equations, i.e.,  one equation for each energy eigenlevel in the energy landscape.  However, solving this many equations is not typically a problem since Eq.~(\ref{eqnEOM}) represents a system of ordinary, first order differential equations, and the coefficients constitute a sparse matrix because most of the probabilities are effectively zero at any given instant of time. These characteristics make the problem mathematically much more tractable than classical moving boundary problems (which are inherently multi-dimensional, partial differential equations). To solve Eq.~(\ref{eqnEOM}) for the present problem, we used a standard MatLab numerical solver function, which typically required less than 30 minutes on a standard laptop computer.}

To solve the system of equations represented by Eq.~(\ref{eqnEOM}), an initial condition defined by a distribution of occupied energy levels is needed.  Eigenstructure information was used to generate initial microstructures for the three types of microstructural evolution considered (sintering, coarsening, and grain growth). For sintering, the initial microstructure was chosen to be an agglomeration of un-sintered particles, for coarsening, it was a distribution of small precipitates within a single grain, and for grain growth the initial microstructure was a collection of small grains within a contiguous solid without voids. Partially canonical distributions along with a perturbation function are then used to calculate the initial probability distributions needed for the SEAQT equation of motion. The partially canonical probabilities of the initial condition, the $p_j^{pe}$, are calculated from,

%VERSION WITH OLD VARIABLES
%\begin{align}
%p_j^{pe}=\frac {{\delta}_j g_j \text{exp}(-\beta^{\text{pe}} \epsilon_j)}{\underset {j} {{\sum}}\phantom{l}{\delta}_j g_j \text{exp}(-\beta^{\text{pe}} \epsilon_j)}
%\end{align}
\begin{align}
p_j^{\text{pe}}=\frac {{\delta}_j g_j \exp(-\beta^{\text{pe}} E_j)}{\underset {j} {{\sum}}\phantom{l}{\delta}_j g_j \exp(-\beta^{\text{pe}} E_j)}
\end{align}
where ${\delta}_j$ takes a value of 1 or 0 depending on whether or not it is assumed that the $j^{th}$ energy eigenlevel is initially occupied or not and $g_j$ and $E_j$ are the degeneracy and energy eigenvalue of the $j^{th}$ eigenlevel. In this equation, $\beta^{\text{pe}}$ is an unknown determined by adding an energy constraint to the system of equations for the $p_j^{pe}$. Once the initial $p_j^{pe}$ are known, an initial non-equilibrium distribution (i.e., initial state) is found using a perturbation function that utilizes the partially canonical probabilities and those of a corresponding canonical distribution.
  
\subsection{Linking State Space to Microstructure \label{State2Microstructure_sec:level3}}

A distinguishing feature of the SEAQT framework is that it works in state space, i.e., Hilbert space or Foch space. The kinetic path is calculated from the component of the entropy gradient perpendicular to the manifold that conserves the generator of motion (e.g., the Hamiltonian and the identity operators). Consequently, it does not depend upon an actual mechanism or even a microstructure to determine how the system evolves. However, to extend its usefulness and help validate the SEAQT framework, the kinetic path information in state space must be connected to the physical microstructures of the evolving material. This is challenging because the degeneracies of some energy levels can be beyond enormous (Figure \ref{fig:DOS} indicates there are more than $10^{1300}$ configurations for the most degenerate levels of the present energy landscape!) and the microstructures corresponding to a single energy level can be quite different from each other. 

This situation poses two problems. The first is that it is impossible to store all the representative microstructures from such an astronomically large collection of possibilities, and the second is that even if one could, randomly selected microstructures along the smooth kinetic path would not necessarily be at all related to each other in time. However, it is possible to select representative microstructures that are consistent with a smooth evolution of microstructure by introducing one or more microstructural parameters in the description of the states. These descriptors can be used to select from among the many degenerate configurations only those that are consistent with a given initial state's evolution to some final stable equilibrium state. Each microstructural descriptor is chosen to reflect an important physical characteristic.  For example, relative density, grain boundary length, and surface length are appropriate descriptors to track sintering kinetics.  In the cases of precipitate coarsening and grain growth, appropriate descriptors are the average precipitate size (area) and grain size (area), respectively.

Once an appropriate set of descriptors are selected, state space is linked to the system's microstructural evolution with the following procedure.  First, the Replica Exchange Wang-Landau algorithm is run to establish the energy levels and the density of states for the energy landscape. At the same time, the values for one or more microstructure descriptors are calculated and recorded for 
%a selected fraction of 
the times an energy level is visited by the algorithm.
%\textcolor{OliveGreen}{ TO REMOVE (e.g., about 20\% of the visits)}. 
Each energy level is characterized by arithmetically averaging the descriptor(s) over the recorded times of visits to the level. Second, an initial state on the energy landscape is selected and the SEAQT equation of motion is solved to find the kinetic path through state space. The energy levels along this path with non-zero occupation probabilities typically represent a very small subset of all the available energy levels; they are the only levels for which microstructure information need be stored. Third, the Replica Exchange Wang-Landau code is re-run to record representative microstructures only for this small subset of energy levels and that also have microstructural descriptor values close to the averaged value for the energy level. This subset of representative microstructures is indexed by energy level and one or more arithmetically-averaged descriptor value(s). 

Lastly, at each moment of time along the SEAQT kinetic path, the occupation probabilities are used to calculate the system energy (an expectation value) and weight-averaged descriptor values.  These values at each time are used to select the closest microstructure from the collection of stored representatives. The resulting time sequence of microstructures evolves smoothly and continuously along the kinetic path through state space. Additional properties such as the relative density and precipitate size distribution of the system microstructure can, of course, also be tracked in time. 

To scale the results predicted by the SEAQT equation of motion to real time, the relaxation parameter, $\tau$, from Equation (\ref{eqnEOM}) can be linked to experimental data or to some dynamical model of the phenomena involved. The former approach is used here. Thus, $\tau$ as a function of the real time, $t$, is determined from experimental data found in the literature. 

Finally, grain area calculations for the individual energy levels are based upon the percolation algorithm found in \cite{Hoshen1976}. The algorithm functions by assigning tags to individual pixels in the modeled domain. When particles of like orientation are checked the system assigns the smallest available tag to both grains and adds the counted area of the grains to this tag. This function allows for efficient calculation of grain size in a single pass of a given lattice. 

\textcolor{Black}{The grain boundary length was calculated as the sum of the lengths of the pixel edges making up grain boundaries in the system. The initial value was adjusted to match the minimum grain size of the experimental data used to test the model.}

\textcolor{Black}{Additionally, the precipitate and grain sizes were calculated by assuming they are approximately circular in shape.  The grain (or precipitate) area was easily tracked in the model by simply summing the area of the pixels with $q$-spins greater than zero, and this area was then used to approximate the grain size from the relationship, $r = \sqrt{A/{\pi}}$.}

\section{Results}

\subsection{Sintering}
\label{Results_subsec:Sintering}

\begin{figure*}[h!]
\begin{center}
\includegraphics[width=.7\textwidth]{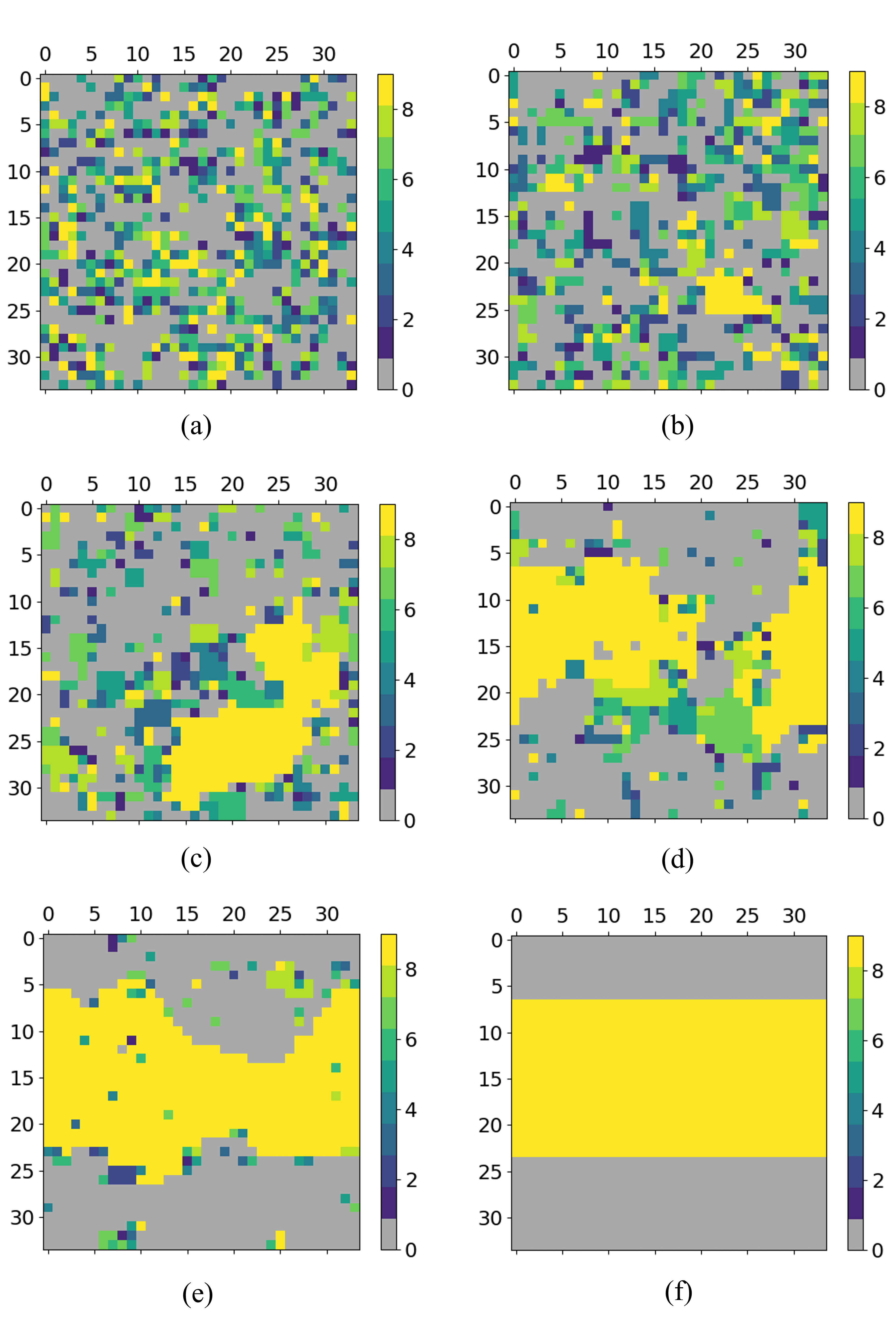}
\caption{Sequence of representative microstructures during sintering. Each image represents a weighted average of the expected states obtained from the SEAQT equation of motion and a descriptor that links the energy levels to the microstructure. The descriptor in this case is the average grain size. Panel a) is the initial state and panels b) through f) are example microstructures along the kinetic path to stable equilibrium. }
\label{SinteringSequence}
\end{center}
\end{figure*}

The evolution of microstructure during sintering is shown in Figure \ref{SinteringSequence}, which presents a time series of microstructures of the material subsystem (the thermodynamic system is a composite system of the material plus the thermal reservoir of Equation \ref{eqnEOM}). Part a) of the figure is the initial microstructure. Each pixel in the figure represents a powder particle 10 nm on a side. The pixel colors indicate crystal orientations, and gray pixels are voids. Different colors adjacent to each other indicate either two grains separated by a grain boundary or a grain with a surface. The maximum number of possible orientations is 50. The physical parameters chosen for the simulation correspond to sintered zirconia with a surface energy of 2.570 J/m$^2$ and grain boundary energy of 0.987 J/m$^2$ \cite{Trunec2007, Tsoga1996}. 

Initially, the energy of the material subsystem is distributed over a narrow set of energy levels associated with many small powder particles. As the material subsystem in Figure \ref{SinteringSequence} moves toward stable equilibrium, the steepest entropy ascent principle distributes the subsystem energy more uniformly over the available energy levels. Since the energy of the subsystem in this model arises only from surface and grain boundaries \footnote{there is no thermal vibrational contribution from the solid in the subsystem energy given by Equation (\ref{TotalEnergy}). However, a harmonic oscillator term could be added if the actual temperature of the solid is an important consideration.}, the removal boundaries during sintering is accomplished by transferring heat from the material subsystem to the thermal reservoir.  When the evolution of states predicted by SEAQT equation of motion are converted to microstructures, the common physical features of sintering and grain growth are evident. For example, the initial small, single-crystal powder particles agglomerate to form larger polycrystalline particles with necks between them (Figure \ref{SinteringSequence}b), and these polycrystalline particles gradually grow in size (Figures \ref{SinteringSequence}c--e). The smaller powder particles eventually disappear entirely as a single solid mass appears (Figure \ref{SinteringSequence}f). Also, the grains within each of the polycrystalline particles grow in size during the process (Figure \ref{SinteringSequence}b--d).   Within the larger particles, one grain orientation eventually grows at the expense of all the others, and at stable equilibrium, the entire solid becomes a single-crystal with minimum surface area (the flat surfaces in Figure \ref{SinteringSequence}f result from periodic boundary conditions on the simulation domain).

\begin{figure*}[h!]
	\begin{center}
		\includegraphics[width=1\textwidth]{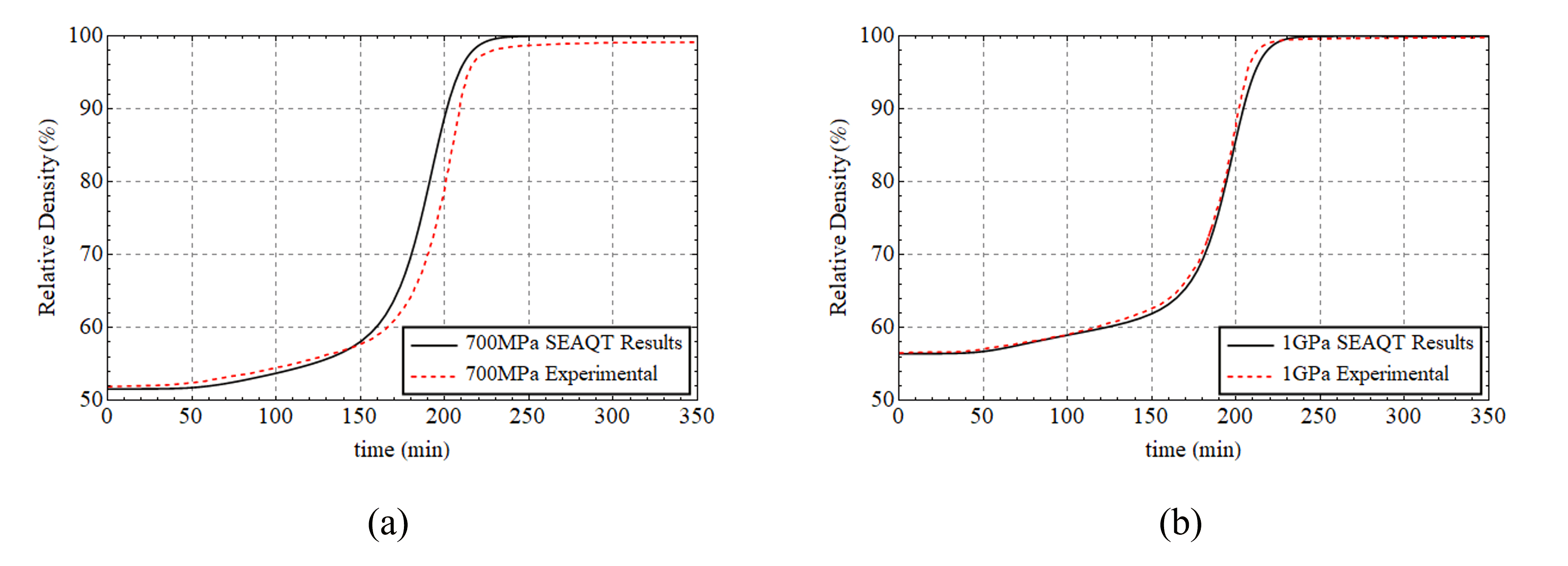}
		\caption{The relative density (a microstructural descriptor linked to energy levels) during sintering calculated from the SEAQT equation of motion. \textcolor{Black}{Figures (a) and (b) compare the predicted and experimental density of zirconia powder during pressureless isothermal sintering at $1100\;^{\circ}$C after compacting with a pressure of 700 MPa and 1000 MPa, respectively \cite{Trunec2007}.} The differences in compaction pressures affect the starting relative density values and are modeled by averaging the relative density for 3000+ states at two separate initial probability evolution  conditions }
		\label{SinteringRelativeD}
	\end{center}
\end{figure*}

\begin{figure*}[h!]
	\begin{center}
		\includegraphics[width=.6\textwidth]{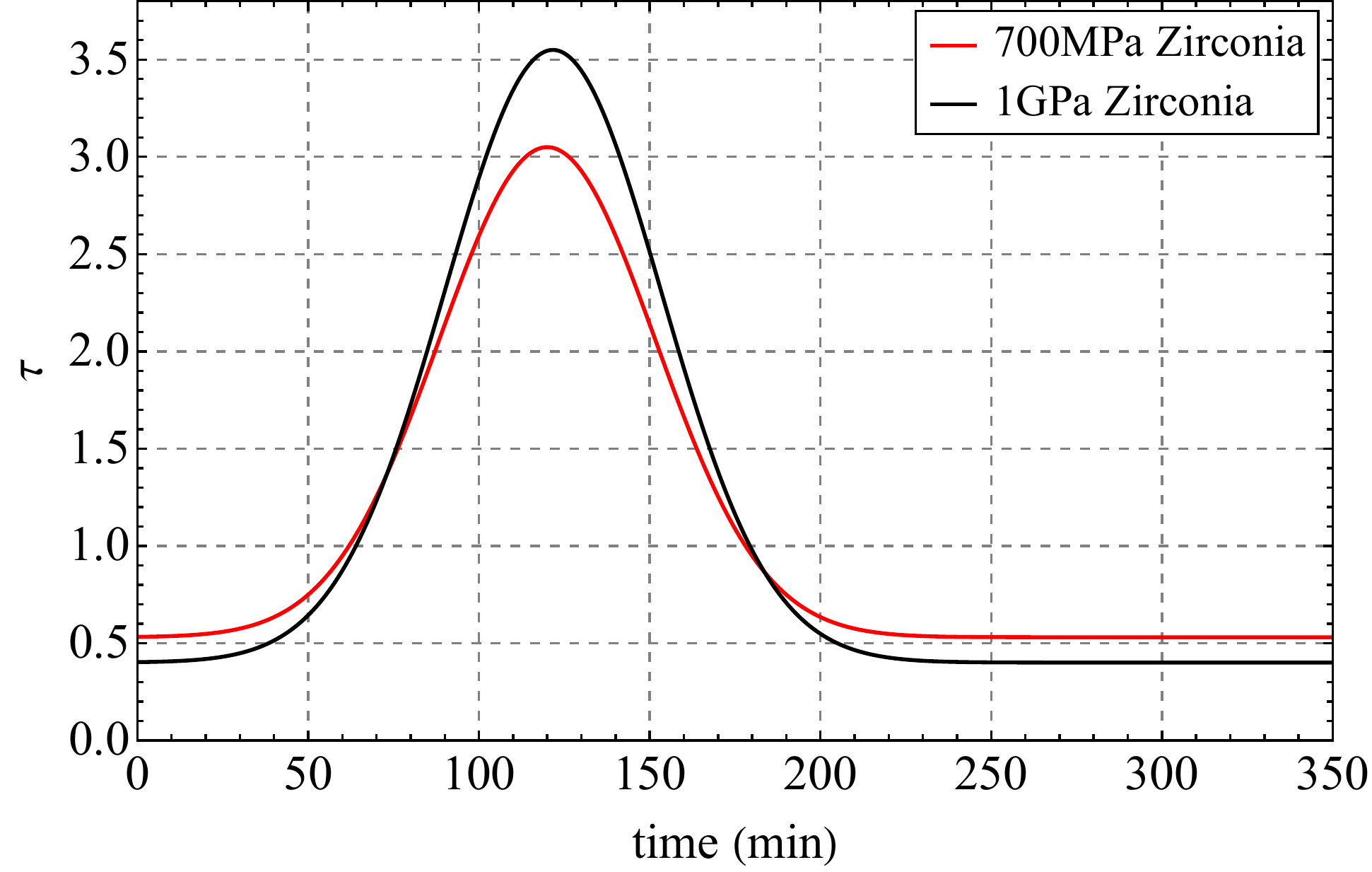}
		\caption{Scaling factor $\tau$ versus real time in minutes. The variance in the plotted values for $\tau$ effect the rapidity of certain phases in the microstructural evolution in the real system versus the initial simulation results}
		\label{TauVsTime}
	\end{center}
\end{figure*}

The change in relative density associated with this microstructural evolution is shown in Figure \ref{SinteringRelativeD}. The relative density descriptor is calculated from averaged configurations of the thermodynamic states represented by the probability distributions predicted by the SEAQT equation of motion. The predicted relative density in Figure \ref{SinteringRelativeD} deviates only slightly from the experimental results of \cite{Trunec2007}. The largest deviation, which occurs at later sintering times for zirconia compacted at 700 MPa (Figure \ref{SinteringRelativeD}a), suggests equilibrium was not reached because the experimental relative density failed to reach 100$\%$.

Note that the SEAQT framework tracks system evolution through state space rather than through a microstructural space as is done with approaches like kinetic Monte Carlo (KMC). This creates important differences between the two methods, notably with regard to how representative configurations of the microstructure  are selected and evolve. In KMC models, individual microstructures or snapshots are sequentially linked in time and are used to approximate material processes. Many KMC models also commonly begin with an idealized starting structure and may utilize algorithms to reduce the computational complexity {\cite{Braginsky2005,Bjork2013}}. In contrast, the SEAQT framework utilizes state-based properties to track the kinetic path so that the selected microstructures are linked by evolving properties, i.e., energy and entropy, versus being explicitly linked in time. In addition, to maintain the generality of the model within the domain of the system, unconstrained particle exchange is allowed. 

This difference means that greater microstructural variety may be present for a given state in the SEAQT framework, and the average morphology for a given energy level has the potential to differ significantly from a similar level visited in a KMC simulation. In other words, the average state-based morphology in the SEAQT framework possesses a greater number of similar permutations than a similar state visited in a KMC model. Thus, under KMC, a specific morphology may exert an inordinate influence on the state's canonical properties. In addition, averaged state properties, such as grain size and particle size distribution, may vary for a KMC simulation, which visits the same energy level. This sampling problem makes it necessary to statistically average multiple KMC runs to obtain representative properties. This sampling problem is avoided in the SEAQT framework because all the evolving properties are expressed as time-dependent expected values.

\textcolor{Black}{The 2-dimensional relative density is the ratio of the area occupied by non-void pixels to the total pixel area. In Monte Carlo sintering simulations~\cite{Braginsky2005,Bjork2013,Zhang2019}, this is commonly calculated from the number of occupied and vacant lattice sites within a given simulation region, which is typically taken from the interior of the coarsening system. In our state-based approach, sintering agglomeration can take place anywhere and is not constrained to any particular region. For this reason, relative density was calculated in this work by eliminating the unattached particles in each simulation configuration and calculating the fraction of solid sites (non-zero Potts $q$-spins) from the remaining regions of agglomerating particles. This procedure yielded a relative density directly comparable to experimental densities as shown in Figure~\ref{SinteringRelativeD}.}

Overall, the experimental kinetics are described closely by the steepest entropy ascent principle. The predicted relative density in Figure \ref{SinteringRelativeD} has an asymmetric S-shaped curve that reflects changing stages in the state of the subsystem and as well as the underlying microstructure evolution. The initial stage before significant grain growth corresponds to an initial incubation-like period of slow particle consolidation. Afterwards, the material transitions to a rapid densification stage with concurrent grain growth. This is followed by a final asymptotic stage during which the rate of property evolution decreases significantly. Microstructurally, this stage is characterized by a reorientation and agglomeration of the largest grains and a steady reduction of the remaining individual grains in the subsystem. As already mentioned above, these results correspond closely with yttria-stabilized zirconia sintering results used for comparison \cite{Trunec2007}. The relative density increase exhibits similar beginning and end lags and an intermediate stage of significant growth. However, the final stage of the experimental results for the yttria-stabilized zirconia still contains multiple grain boundaries not present in the SEAQT results. This can be attributed to the conformational freedom of the model; no constraints were placed on the model to prevent it from achieving stable (single-crystal) equilibrium.

The dissipation parameter, $\tau$, in the SEAQT equation of motion can be adjusted to fit predicted kinetics to experimental data.  The predicted and experimental density of zirconia powder during pressureless isothermal sintering at $1100^{\circ}$C after compacting at a pressure of 700MPa and at 1000MPa~\cite{Trunec2007} are compared in Figure~\ref{SinteringRelativeD}). The time-dependent  $\tau$ functions used to match these two sets of experimental data are shown in Figure~\ref{TauVsTime}. Physically, $\tau$ reflects how fast the system moves along the kinetic path in state space. 

\begin{figure*}[h!]
	\begin{center}
		\includegraphics[width=0.45\textwidth]{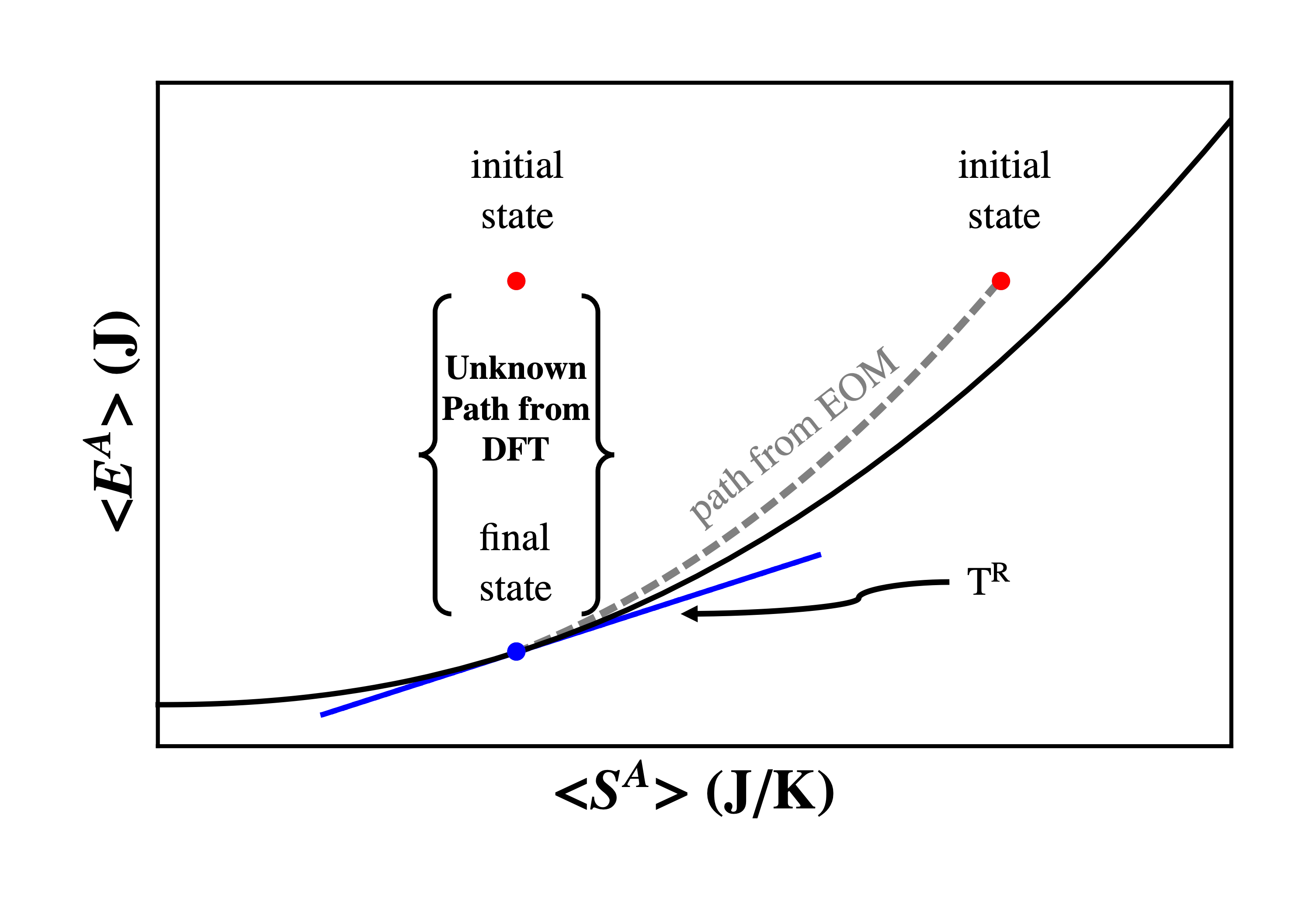}
		\vspace{0.5truecm}
		\includegraphics[width=0.44\textwidth]{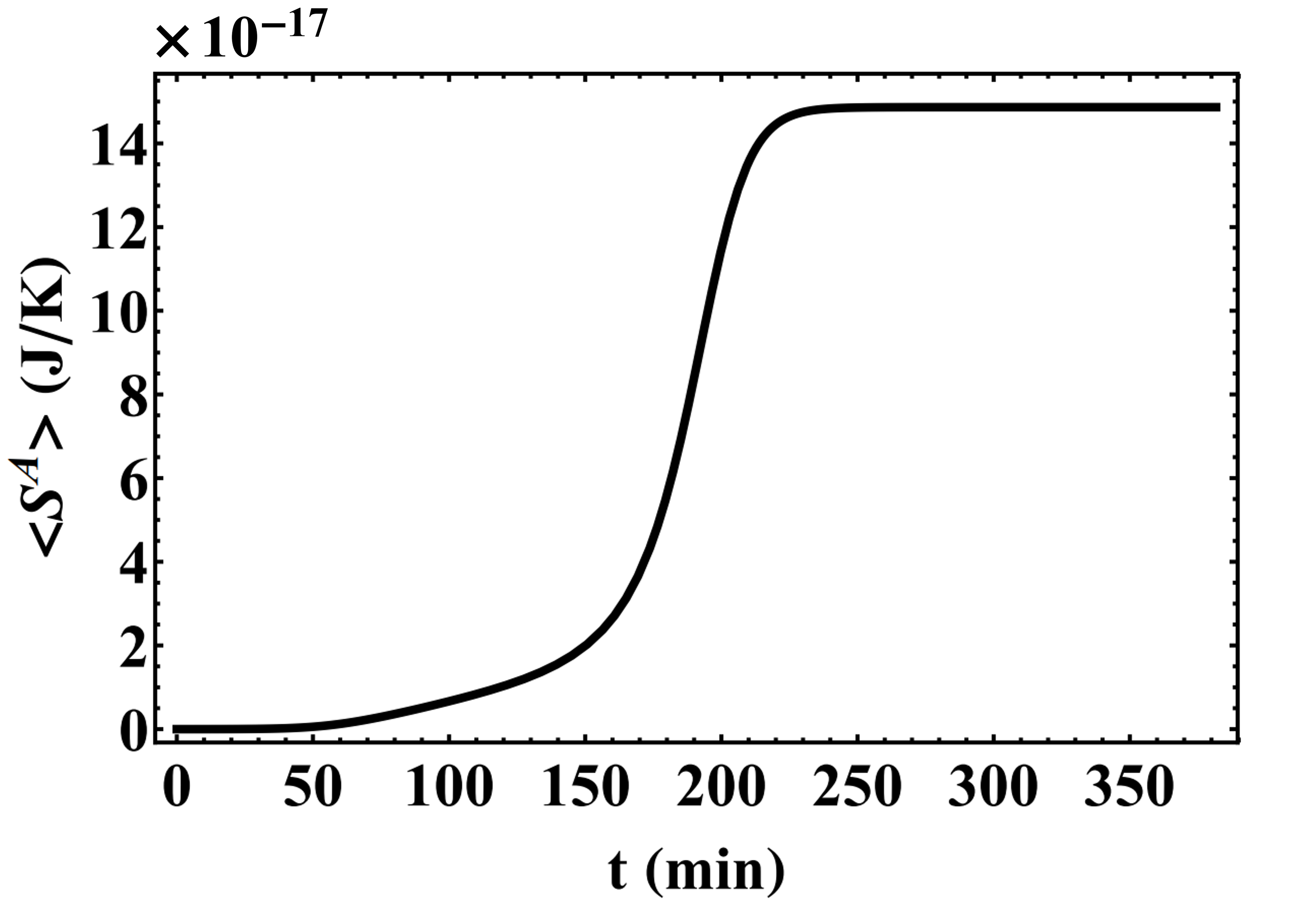}
		\caption{(Left) A schematic plot of the energy versus entropy of the material subsystem.  The bounding solid curve represents the set of equilibrium states and the two red points are possible non-equilibrium states. The SEAQT equation of motion provides a unique path from an initial state (red point on the right) to stable equilibrium along the dashed gray curve. Techniques like density functional theory and kinetic Monte Carlo methods minimize energy without regard to energy dissipation so there is no thermodynamic path between an initial state and final equilibrium. (Right) The entropy production that drives the subsystem from a non-equilibrium initial state to equilibrium along the kinetic path determined by the SEAQT equation of motion. Entropy of the composite system increases despite the fact that heat transferred out of the subsystem into the reservoir reduces the entropy of the material subsystem.}
		\label{E-Sinfo}
	\end{center}
\end{figure*}

Now, one of the distinct advantages of working in state space is that the steepest-entropy-ascent (or, equivalently, maximum entropy) principle is able to identify the unique kinetic path a system follows without any prior knowledge of the physical mechanisms involved. This is illustrated schematically in Figure \ref{E-Sinfo}a which represents a plot of the energy of the material subsystem as a function of its entropy. The solid bounding curve represents the set of equilibrium states; the ground state is the energy at the point $S=0$. The two red points represent arbitrary non-equilibrium states. % note that temperature is undefined at these two points because they lie away from the equilibrium curve where temperature is formally defined as $T \equiv \frac{d E}{d S}$.
The information contained in the SEAQT equation of motion provides the unique path from an initial state to stable equilibrium (the dashed gray curve) that maximizes entropy ascent at each point in time. In the sintering case, the entropy of the material subsystem {\em decreases} from the initial state as surface and grain boundaries are removed from the solid and heat is transferred from the subsystem to the reservoir. However, the net entropy of the overall composite system (subsystem plus reservoir) increases along the path to equilibrium (see Figure \ref{E-Sinfo}b), and this entropy ascent spontaneously drives the sintering process. It is worth noting that most computational tools for finding stable states --- like density functional theory, kinetic Monte Carlo methods, and molecular dynamics ---  all minimize system energy without regard to how energy is dissipated. Thus, these tools cannot predict the thermodynamic path between an arbitrary initial state and a stable equilibrium state. Perhaps most surprisingly, the SEAQT equation of motion predicts the microstructural evolution sequence (Figure \ref{SinteringSequence}) without any explicit assumptions about how the surfaces and grain boundaries physically behave during the sintering process.  

\subsection{Precipitate Coarsening \label{Results_subsec:Ostwald}}

Appropriate regions of the energy landscape (Section \ref{REWL_sec:level2} and Figure~\ref{fig:DOS}) can be used to describe precipitate coarsening kinetics.  Specifically, we represent a precipitate phase as one  non-zero $q$-spin of the Potts model (or one color) contained within a parent phase, or matrix, which is designated by another non-zero $q$-spin (a second color).  By choosing an initial state in a region of the energy landscape with only two spins and no free surfaces, the distribution of spins can represent precipitate particles undergoing coarsening within a matrix phase.

\begin{figure*}[h!]
\begin{center}
\includegraphics[width=.7\textwidth]{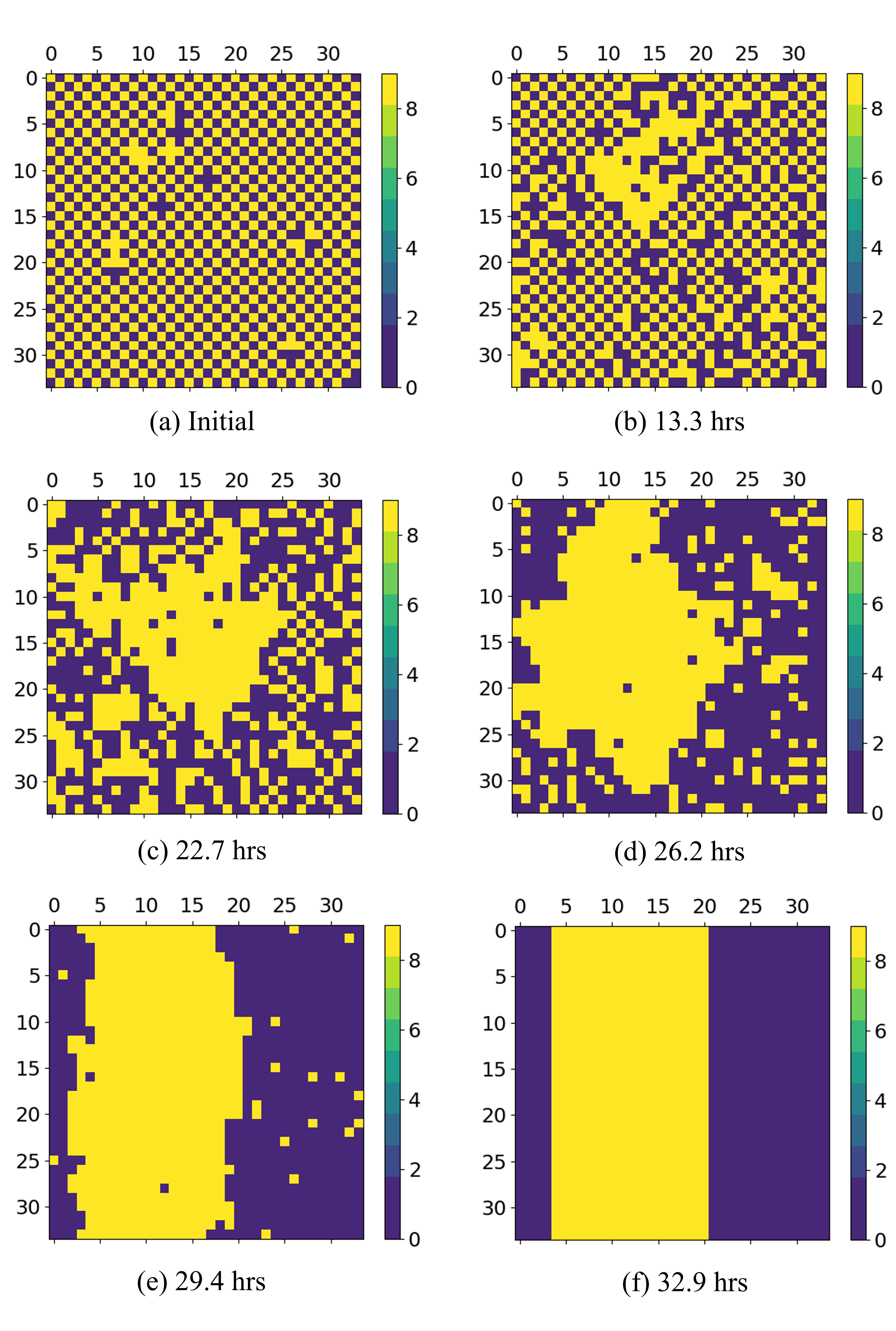}
\caption{Sequence of representative microstructures during coarsening of Li-rich $\delta'$ precipitates (yellow) in an aluminum matrix (purple). Each image represents a weighted average of the expected states obtained from the SEAQT equation of motion and a descriptor that links the energy levels to the microstructure. }
\label{OstwaldSequence}
\end{center}
\end{figure*}

The SEAQT predicted coarsening kinetics of precipitates are compared against experimental data for the coarsening of Al$_{3}$Li precipitates (designated $\delta'$) in an Al--Li solid solution (the $\alpha$ matrix) \cite{Williams1975, Noble1999, Jensrud1984, Pletcher2012a}. This alloy system is convenient because $\delta'$ precipitates are nearly spherical and are thus comparable to the morphologies expected from the isotropic boundaries assumed in the energy landscape.  The representative microstructures calculated during precipitate coarsening are shown in Figure~\ref{OstwaldSequence}; the initial state is Figure~\ref{OstwaldSequence}(a).

\begin{figure*}[h!]
\begin{center}
\includegraphics[width=1\textwidth]{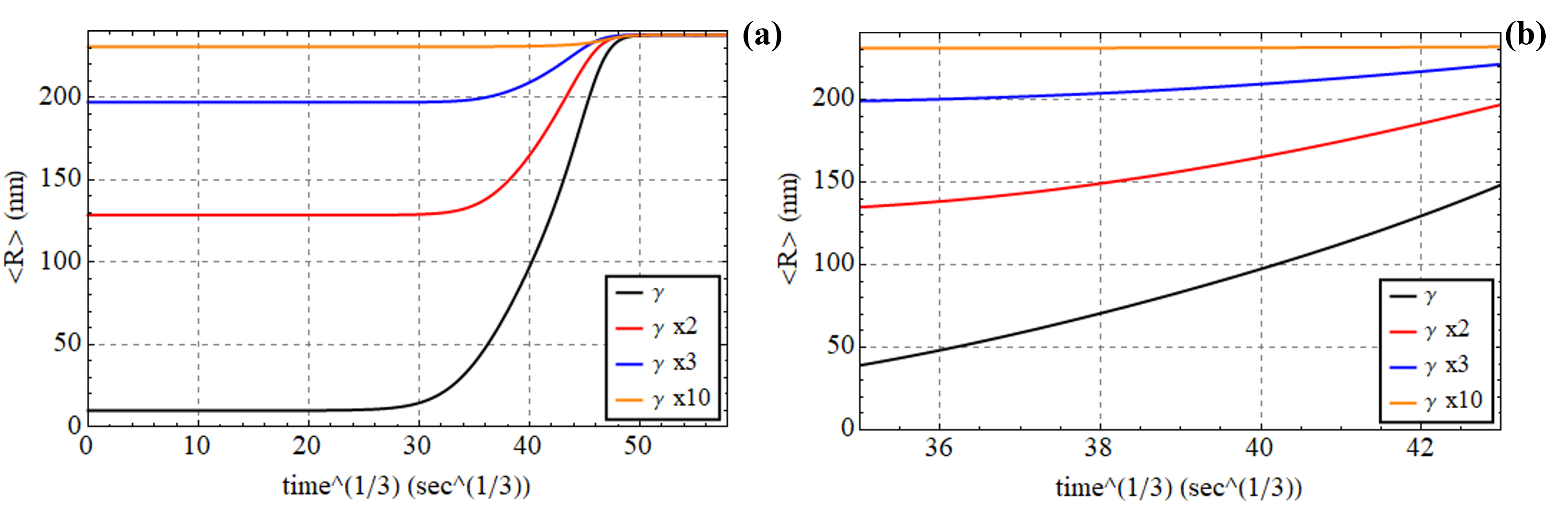}
\caption{\textcolor{Black}{Expected radius of Li-rich $\delta'$ precipitates during coarsening predicted by the SEAQT equation of motion: (a) a range of times from an initial state of many small precipitates to stable equilibrium, and (b) a selected intermediate time period over which $t^{1/3}$ kinetics are roughly linear.  The four curves in each figure represent coarsening kinetics for $\gamma = 0.005 \,{\text J}/{\text m^2}$ (black curves), and 2, 3 and 10 times this boundary energy (red, blue, and orange curves, respectively).} }
\label{CoarseningComparison}
\end{center}
\end{figure*}

\begin{figure*}[h!]
\begin{center}
\includegraphics[width=.8\textwidth]{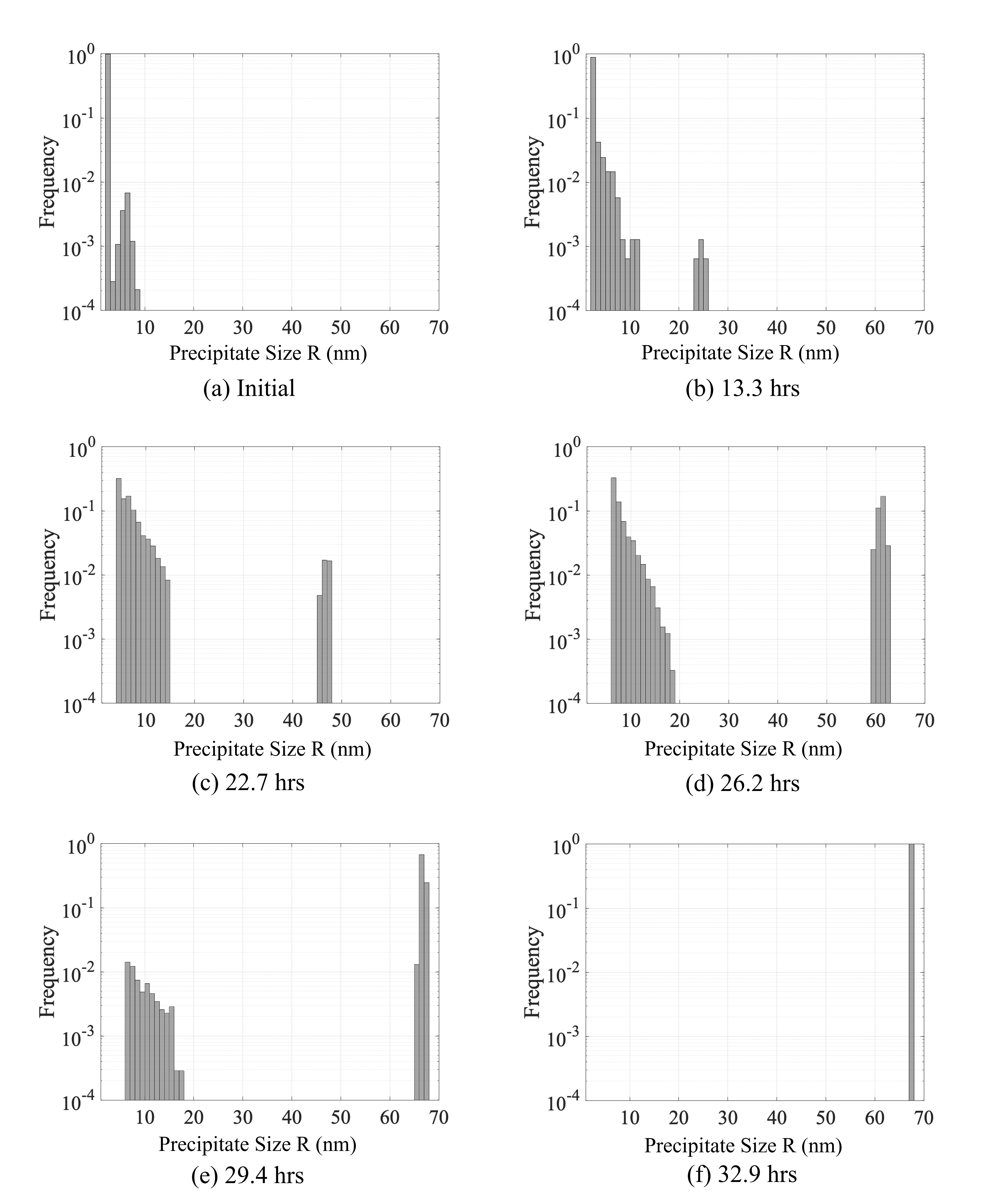}
\caption{The SEAQT-predicted relative frequency of Li-rich $\delta'$ precipitates in an aluminum matrix at different stages of coarsening: (a) the initial distribution, (b) after annealing 13.3 hrs at $225^{\circ}$C, 
(c) after annealing 22.7 hrs at $225^{\circ}$C, (d) after annealing 26.2 hrs at $225^{\circ}$C, (e) after annealing 29.4 hrs at $225^{\circ}$C, and (f) after annealing 32.9 hrs at $225^{\circ}$C. The vertical axis in each panel represents the frequency of precipitate sizes averaged over multiple representative lattices for each energy level. A log scale is utilized to reveal the small frequencies.}
\label{OstwaldSequencePSD}
\end{center}
\end{figure*}

In the actual alloy, the initial state for the coarsening process is a dispersion of small precipitates produced by nucleating $\delta'$ precipitates within a supersaturated grain of $\alpha$. Coarsening takes place during annealing at  at $225^{\circ}$C~\cite{Williams1975,Noble1999,Jensrud1984,Pletcher2012a}.  Only one $\alpha$ grain is considered, so there is only one matrix orientation. The energetics of nucleation in this system ensures that only one $\delta'$ orientation appears within any particular $\alpha$ grain, so the precipitate coarsening kinetics can be described with only two colors separated by an interphase boundary between the $\alpha$ and $\delta'$ phases. In the model microstructure (Figure \ref{OstwaldSequence}), the $\alpha$ phase is represented by the purple pixels and the $\delta'$ phase by the yellow pixels.  The $\delta':\alpha$ boundary energy, $\gamma$ is quite small in the Al--Li system~\cite{Hoyt1991}; it is assumed to be 0.005~${\text J}/{\text m^2}$ in the SEAQT coarsening simulation. 

An increase in the average size of precipitates is evident in the simulation microstructures of Figure~\ref{OstwaldSequence}. The elongated precipitate with planar boundaries that appears in the longest two times is a consequence of periodic boundary conditions imposed on the simulation domain.  Although the coarsening path through state space is not tracking a particular microstructure configuration, the SEAQT framework is able to capture the approximately circular growth of the $\delta'$ precipitates (Figure \ref{OstwaldSequence}(b) and (c), the disappearance of small precipitates (Figure \ref{OstwaldSequence}(d) and (e), and the eventual dominance of large ones (Figure \ref{OstwaldSequence}(e) and (f).

\textcolor{Black}{Analytical models for coarsening in 2-dimensions~\cite{Marqusee1984} suggest the mean precipitate radius should follow a $t^{1/3}$ dependence. Figure~\ref{CoarseningComparison} presents the expected value of the precipitate radius, $R$,  predicted by the SEAQT model as a function of $t^{1/3}$ for four different interphase boundary energies.  In each case, the precipitate radius approximates a $t^{1/3}$ dependence, but only over a very limited range of times. Overall, the SEAQT-predicted radius exhibits an {\em S\/}--shaped time dependence and slows asymptotically as equilibrium is approached. This does not necessarily contradict analytical coarsening models. The absence of a universal $t^{1/3}$ dependence is most likely a consequence of the fact that the SEAQT model is not constrained by assumptions typically made to reach an analytical expression (e.g. small precipitate volume fractions, a mean-field approximation for the average solute distribution, a cutoff precipitate size, etc). The different coarsening rates for the four different boundary energies demonstrate that changes in the energy landscape (even relatively small changes in the energy scale) can alter the steepest-entropy-ascent path and lead to different overall kinetics.}

Coarsening is commonly characterized using precipitate size distributions. The size distributions of the SEAQT simulation are reported here using the precipitate size, $R$, and the log of the frequency of that size. To reduce noise in the statistics arising from the limited domain size, only precipitates larger than 10\% of the largest precipitate were included in the distribution. The SEAQT-simulated precipitate size distributions at the initial state and the final equilibrium state as well as at four intermediate non-equilibrium states along the system's kinetic path are shown in Figure~\ref{OstwaldSequencePSD}. A bimodal size distribution develops as larger precipitates begin to evolve from the initial distribution. The distribution of small precipitates does not change much over the course of the simulation. The radius of the larger precipitates gradually shifts to larger and larger sizes, and eventually ends at a radius of 67 nm (the largest size available with this particular simulation domain). 

\begin{figure*}[h!]
\begin{center}
\includegraphics[width=0.7\textwidth]{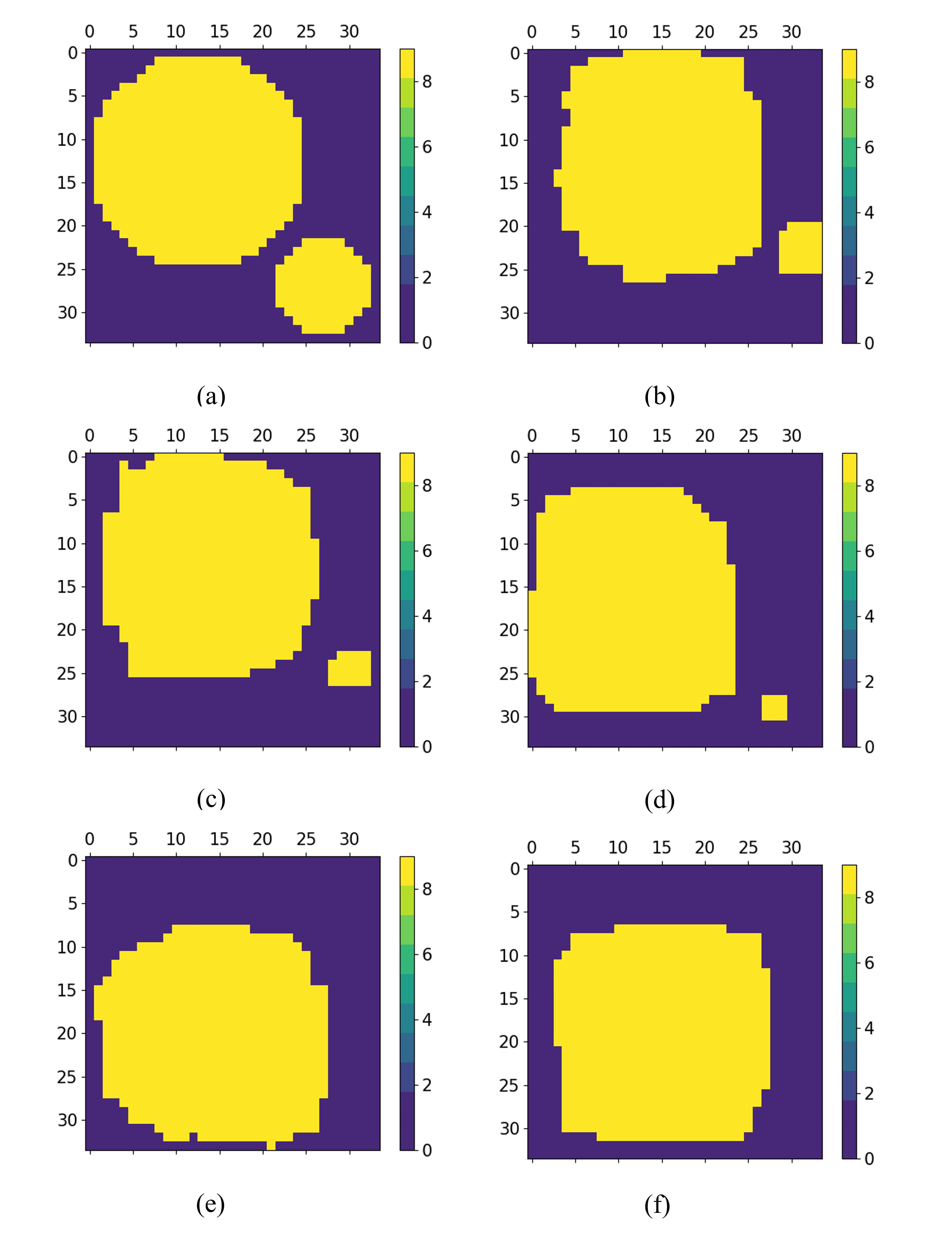}
\caption{\textcolor{Black}{Coarsening of two isolated precipitates.} 
}
\label{TwoPart}
\end{center}
\end{figure*}

\begin{figure*}[h!]
\begin{center}
\includegraphics[width=.7\textwidth]{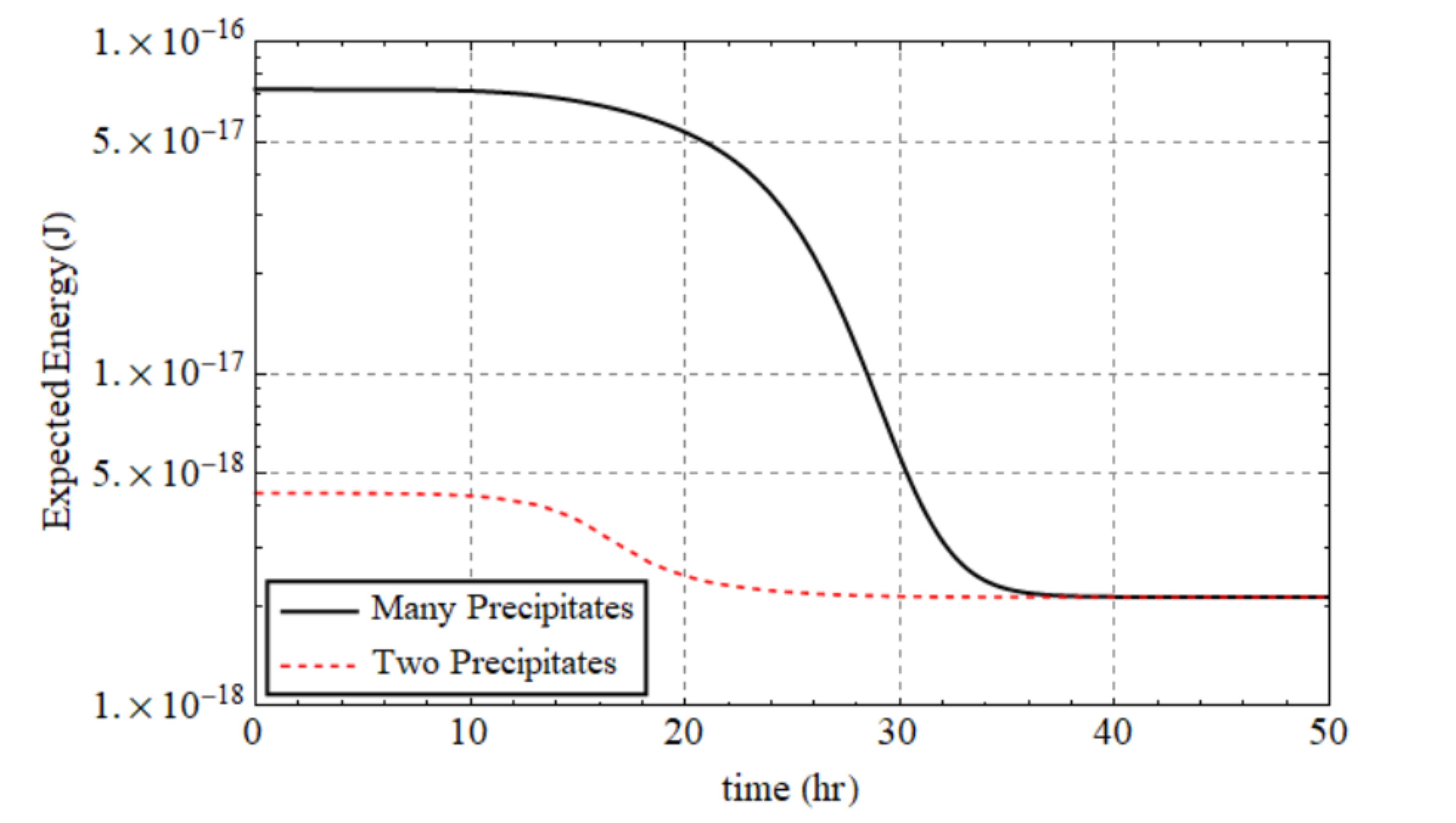}
\caption{\textcolor{Black}{Expected energy as a function of coarsening time. The black curve represents the case of many $\delta'$ precipitates shown in Figure~\ref{OstwaldSequence}, and the dashed red curve corresponds to the two $\delta'$ precipitates shown in Figure~\ref{TwoPart}.} 
}
\label{TwoPartExE}
\end{center}
\end{figure*}

\textcolor{Black}{It is interesting to use the energy landscape to explore coarsening behavior from a very different starting precipitate distribution. Figure~\ref{TwoPart} shows the evolution of two isolated $\delta'$ precipitates. To describe two-particle coarsening mechanistically and compare it with multiple particle coarsening, it would be necessary to reformulate classical Greenwood-Lifshitz-Slyozov-Wagner theory to reflect the new geometry. A feature of the SEAQT framework is that it works from an energy landscape that incorporates all the possible microstructural geometries. For example, Figure~\ref{fig:DOS} includes states consisting of two, three, four, ..., $n$ precipitates and all kinds of different size distributions. Simulating two-particle coarsening simply involves constructing an initial probability distribution associated with several energy levels near this desired microstructural configuration. The set of evolving microstructures in Figure~\ref{TwoPart} was generated from a two-precipitate initial microstructure averaged from three similar energy eigenlevels. The microstructural sequence was obtained with the procedure of Section~\ref{State2Microstructure_sec:level3} using boundary length as the descriptor.  The periodic boundary conditions were relaxed to allow the precipitates to adopt any shape; this choice affects the microstructure evolution, but not the energy landscape or the kinetic path.}

\textcolor{Black}{For a fixed precipitate fraction, the system of two precipitates begins with much less boundary length than the original case shown in Figure~\ref{OstwaldSequence}(a). Nevertheless, both systems must coarsen to the same thermodynamic equilibrium state. Solving the equation of motion to find the time-dependent occupation probabilities in the two cases makes such a comparison straightforward.  The expected energy during coarsening is shown in Figure~\ref{TwoPartExE} (the expected entropy can be calculated in a similar fashion). The black curve represents the expected energy during coarsening of the $\delta'$ precipitates shown in Figure~\ref{OstwaldSequence}, and the dashed red curve corresponds to coarsening of the two $\delta'$ precipitates shown in Figure~\ref{TwoPart}. The initial expected energy of the two-precipitate microstructure, Figure~\ref{TwoPart}(a), is lower than that of the many-precipitate microstructure, Figure~\ref{OstwaldSequence}(a), but the final equilibrium configurations are essentially the same (Figures~\ref{TwoPart}(f) and \ref{OstwaldSequence}(f)). Overall coarsening kinetics are not expected to depend explicitly on the number of precipitates, so the similarity in the way the energy evolves with time in the two cases of Figure~\ref{TwoPartExE} is reasonable.}

\textcolor{black}{While the microstructures and precipitate size distributions predicted by the SEAQT framework during coarsening are qualitatively reasonable (Figures~\ref{OstwaldSequence}--\ref{OstwaldSequencePSD}), they deviate in at least two significant ways from $\delta'$ coarsening in an Al--Li alloy~\cite{Pletcher2012a}.  Most obviously, the limited number of pixels in the simulation make it necessary to represent the initial precipitate microstructure as an over-simplified array of individual squares rather than as a distribution of circles. Also, the use of only first nearest-neighbor interactions in the Potts model biases the energy landscape in a way that shifts the predicted precipitate distribution to smaller  sizes.  Both of these shortcomings can be addressed by computing a more accurate energy landscape with finer energy resolution. However, because the landscape was used for two other applications (sintering in Section~\ref{Results_subsec:Sintering} and grain growth in \ref{Results_subsec:GG}), and these applications do not need as much energy resolution, no additional computational effort was made to refine the energy landscape and improve agreement with experimental coarsening data.}

\subsection{Grain Growth \label{Results_subsec:GG}}

%(Ultimately, we decided to drop the modeling of recrystallization kinetics in favor of nanocrystalline grain growth. This change occurred after looking into the kinetic Monte Carlo and cellular automata modeling of primary recrystallization. Those models utilize an energy associated with the initial deformed or yet-recrystallized region of the material which we do not have in our energy landscape.)

\begin{figure*}[h!]
\begin{center}
\includegraphics[width=.7\textwidth]{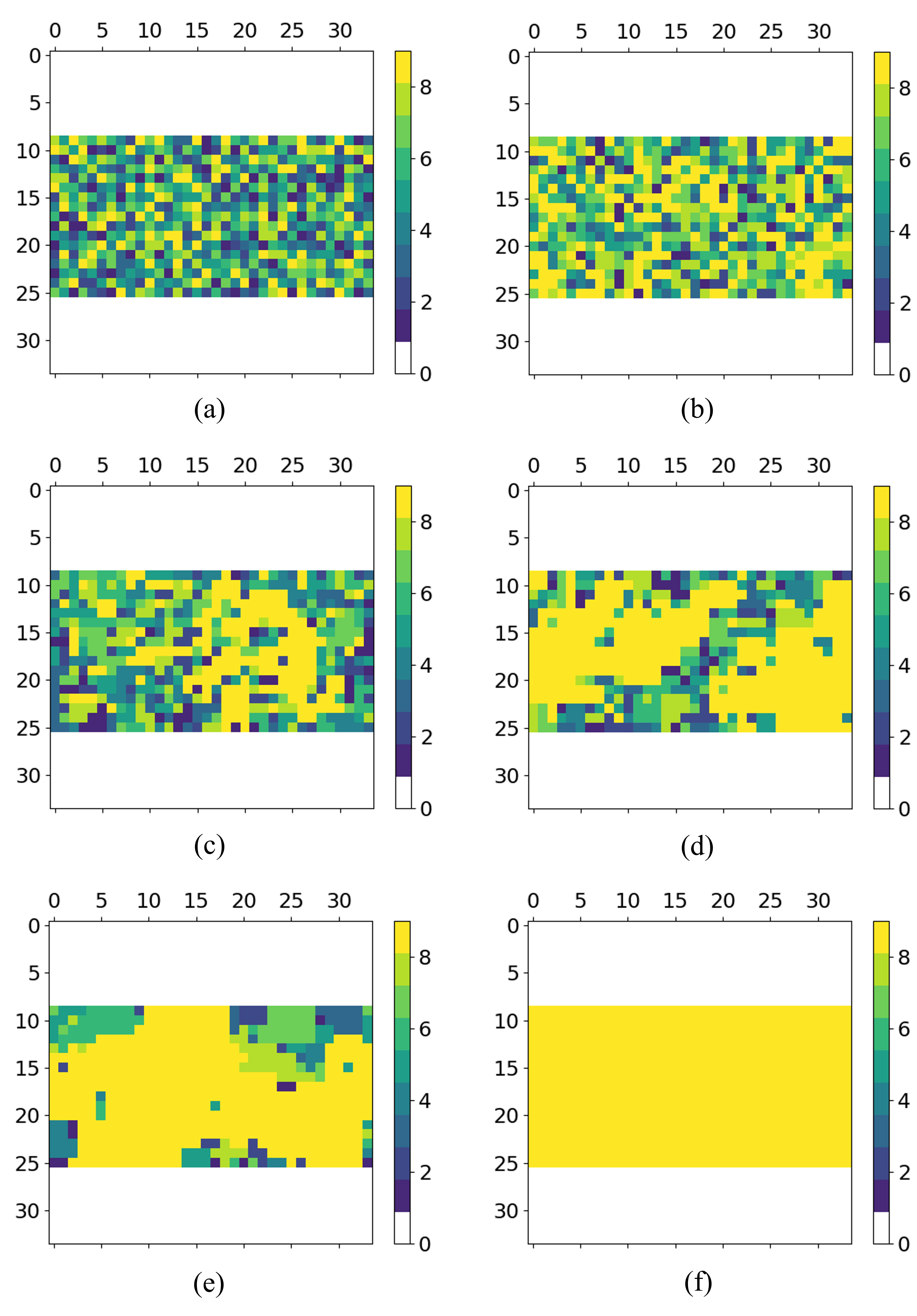}
\caption{Sequence of representative microstructures during grain growth of nanocrystalline Pd. Each image represents a weighted average of the expected states obtained from the SEAQT equation of motion and a descriptor that links the energy levels to the microstructure. The descriptor in this case is the average grain size. Panel (a) shows the microstructure of the initial state of many small grains, panels (b) through (e) provide the microstructures for increasing annealing times, and panel (f) is the microstructure of a single crystal at stable equilibrium. }
\label{GrainGrowthSequence}
\end{center}
\end{figure*}

The kinetics of grain growth predicted by the SEAQT framework from the aforementioned energy landscape (Section \ref{REWL_sec:level2} and Figure~\ref{fig:DOS}) can be simulated by starting with an initial state that represents a fully dense solid with a collection of different grain orientations (different non-zero $q$-spins) represented by different pixel colors. The landscape of Figure~\ref{fig:DOS} included up to maximum of 50 different grain orientations. The representative microstructures predicted by the SEAQT equation of motion are shown in Figure \ref{GrainGrowthSequence} for a sequence of annealing times. The physical size of the initial grains represented by the pixel size was set at 1 nm on a side, which corresponds approximately to the initial grain size in a nanocrystalline Pd system undergoing grain growth at room temperature~\cite{Ames2008}.  The surface boundary energy used for this system is 1.47 J/m$^2$ taken from a weighted average found in Tran \cite{Tran2016}. The grain boundary energy utilized is 0.8 J/m$^2$ \cite{Ames2008}. This system is chosen for its similarities with the microstructural descriptors calculated previously. 

\begin{figure*}[h!]
\begin{center}
\includegraphics[width=.6\textwidth]{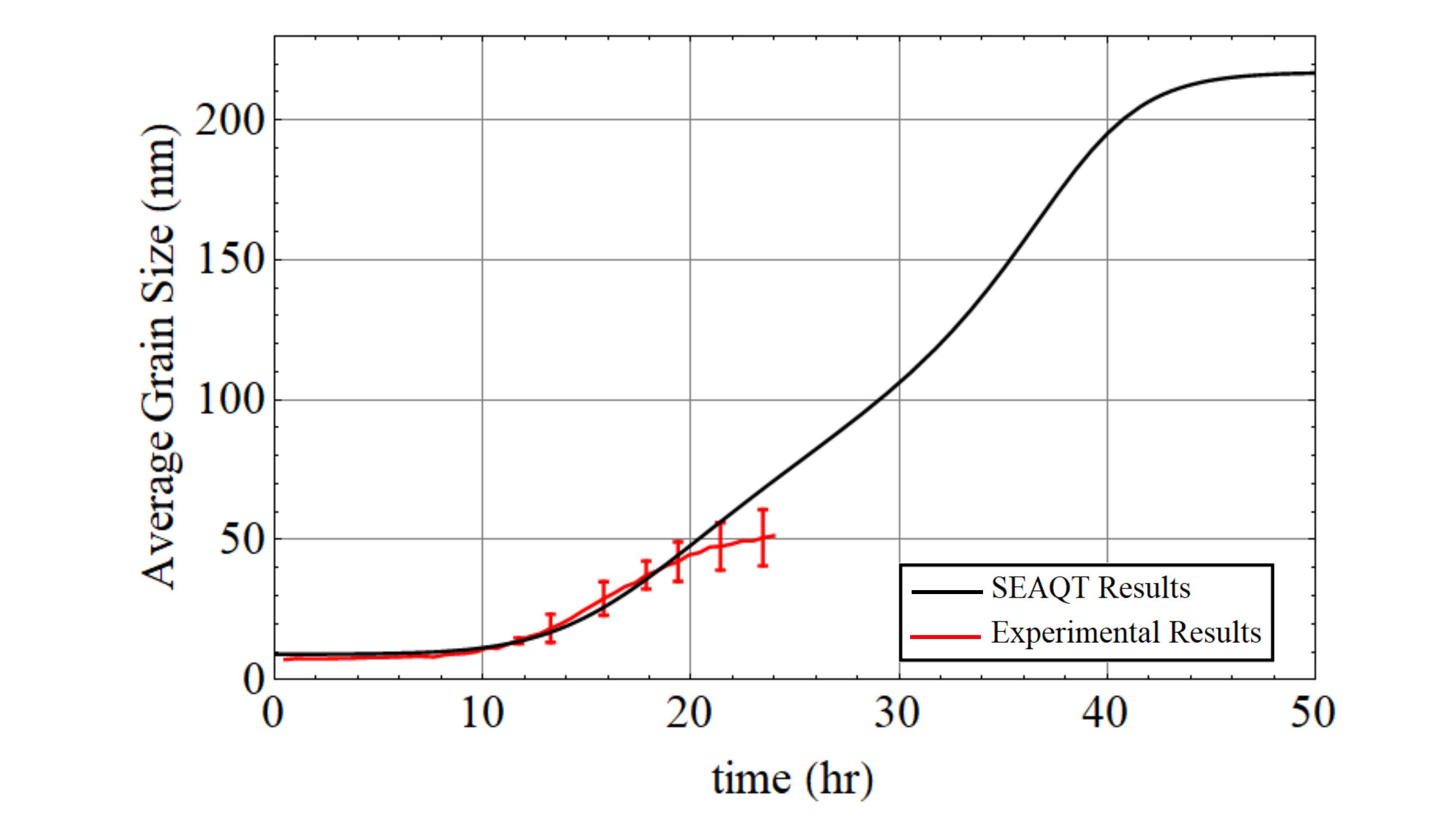}
\caption{\textcolor{Black}{The average grain size of nanocrytalline Pd during grain growth at room temperature. The black curve represents the grain size predicted by the SEAQT equation of motion, and the red curve represents data from reference~\cite{Ames2008}. The predicted curve is extended beyond the largest grain size accessible with the original experiment.} }
\label{GrainGrowthAveGS}
\end{center}
\end{figure*}

The initial state begins with a near-maximum number of grain boundaries, i.e., adjacent pixels all have different colors and, thus, represent grains of different orientations. To consider only grain boundary changes independent of any changes in surface energy, the initial state is a single solid block of individual grains selected to have a minimum number of surface pixels. Because of periodic boundary conditions, this yields planar top and bottom surfaces in Figure \ref{GrainGrowthSequence}. There are no internal voids for this simulation of pure grain growth.  Figure \ref{GrainGrowthSequence}a) is a very different initial condition from that of sintering (Figure \ref{SinteringSequence}a)) and precipitate coarsening (Figure \ref{OstwaldSequence}a)), but each of these simply represent different starting states that can be found on the same energy landscape of Figure \ref{fig:DOS}. 

The microstructural changes predicted by the SEAQT framework in Figure \ref{GrainGrowthSequence} follow the general expectations of grain growth: the average grain size increases in Figures \ref{GrainGrowthSequence}b) through \ref{GrainGrowthSequence}e) as small grains coalesce into larger grains and the larger grains continue to grow at the expense of the smaller grains. At stable equilibrium (Figure \ref{GrainGrowthSequence}f)), the system consists of a single grain with minimum surface area. 

The average grain size (a microstructural parameter linked to the energy levels) and grain size distribution are the descriptors used to characterize grain growth evolution with time. The evolution of the average grain area, a descriptor calculated from averaged configurations of the states represented by the probability distributions, is \text{blue}{compared in Figure \ref{GrainGrowthAveGS} with experimental grain growth data from nanocrystalline Pd \cite{Ames2008}.} The predicted average grain size has an overall ``$S-$shape'' that matches the experimental growth kinetics quite well at short times.  At longer times (beyond annealing times of 20 hours) the experimental data deviates from the predicted kinetics, but there are a couple of possible explanations for this deviation. Grain growth in thin, transmission electron microscopy samples is expected to slow as the grain size approaches the sample thickness (on the order of $100nm$) and thermal grooving begins to pin grain boundary migration.  Ames $et\;al.$ \cite{Ames2008} also note that Pd grain growth begins with an initial slow grain growth followed by a period of rapid abnormal grain growth, and ends with a potential reoccurrence of the initial steady grain growth. The abnormal grain growth violates the statistical self-similarity postulate for a sintering system, which states that evolving systems in time should retain the statistical similarity of their geometric features. Nonetheless, the abnormal behavior is generic in nano-grained systems and may be caused by the release of microstrain as grain size increases \cite{Ames2008}.

\subsection{Discussion \label{GeneralDiscussion}}

The compatibility of the SEAQT framework predictions of microstructural evolutions with those of traditional kinetic models is complicated by a few issues. As stated previously, a representative lattice for a given energy level determined from the Replica Exchange Wang-Landau algorithm, will appear visually distinct from that determined along a given kinetic path of a KMC modeled system. One specific example is the presence of individual pixel grains that remain in the conformational space of the simulation. The presence of these peculiar grains is due to the nature of the Replica Exchange, Wang-Landau process, which estimates the density of states independent of a modeled kinetic path. Often to reach these energy levels, physically unexpected but intuitively easier transitions occur. Transitioning to higher energy levels by placing a grain currently in contact with a larger coarsening mass into a vacant void site is often easier than complex conformations of the existing structure. Multiple similar transitions can lead to individual grains located significantly far apart from the majority of the coarsening mass. Thus, small grains, in proportion to the larger coarsening mass, are partially ignored in the present descriptor calculations such as those for the relative density and precipitate size distribution calculations since their variable locations can nonphysically bias the values output. Future work to counteract this behavior could bias transitions to reduce the probability of particle and vacant site exchanges or restrict translational movement during grain transitions.

Another method to potentially further distinguish between these variable microstructural formations would be the addition of other energetic terms. One example is the addition of $2^{nd}$ nearest neighbor energetic considerations. This addition would allow grain agglomeration to be more easily differentiated energetically because of the higher number of grain sites considered in the energy interaction of a single grain. It would also distinguish lattices with a large number of grains surrounded by vacant sites attributable to the higher energy associated with the transition. This and other additional energetic considerations would result in an increased computational cost from the higher number of energy levels and could potentially limit the size of the simulated system.

The data presented for the precipitate coarsening and grain growth case should be regarded as demonstrating qualitative trends. Although the data aligns closely to expected results, it is intended here to only demonstrate the applicability and flexibility of the SEAQT framework rather than a particular quantitative result. Such a result could be had with a more precise energy landscape. For example, with regard to modeling the precipitate coarsening kinetics, the volume fraction of the precipitate $\delta'$ in the simulated or experimental systems is significantly lower than in the system shown in these articles \cite{Vaithyanathan2000,Williams1975,GrozaBook,Gu1985,Schmitz1992,Pletcher2012a,Pletcher2012b,Hoyt1991}. Commonly simulated values for the volume fraction do not exceed $>30\%$ \cite{Pletcher2012a}. In experimental work the volume fraction trends towards $<12\%$ \cite{GrozaBook}. The constraints on this fraction are either physical and due to solubility limits on the amount of precipitate phase formed or conform to kinetic theory, which relies on a lower density to match experimental results more closely. Thus, the main inadequacy of using the current eigenstructure or landscape with the SEAQT equation of motion to predict precipitate coarsening kinetics is the higher volume fraction of the precipitate phase, which increases the potential for interactions between individual coalesced precipitates. The resultant formations can impede the formation of spherical precipitates and skew the expected size distribution. Improving the calculation of precipitate coarsening kinetics will require calculating a new, more accurate eigenstructure that accounts for precipitate interactions or has a smaller solute concentration. In that case, a new eigenstructure could also be constructed to remove boundary interactions as a free parameter. Removing this degree of freedom in the model and thus, reducing the number of occupied precipitate sites would also allow for the possibility of simulations of larger systems with higher spatial resolution. Additionally, this would make a more efficient estimation of the density of states possible due to the reduction in the number of available energy levels.

Another limitation of the present energy landscape or eigenstructure is its limited spatial resolution. The mass conservation constraint used in the model in Section \ref{Results_subsec:GG} forces the modeling of grain growth behavior to occur along a path of minimum surface boundaries. This is intended to mimic grain growth in a system with a defined number of vacant sites. However, because of the lack of spatial resolution for this particular landscape, the dimensional configurations are effectively limited to a single spatial direction. This can prevent the formation of spherical grains and bias the particle size distribution.

A more robust method of calculating grain growth would require a new eigenstructure where grain boundary energy is the only energetic parameter. Removing the degree of freedom of surface boundaries, as is done in the precipitate coarsening case, allows for the calculation of a larger number of available sites. Additionally, other types of kinetics, like recrystallization, could be modeled by including additional energy terms for stored plastic deformation in the energy eigenstructure.

Finally, it is important to note that most of the computational resources required by the SEAQT framework are needed to generate the energy landscape (Figure \ref{fig:DOS}) via the Replica Exchange, Wang-Landau algorithm.  This algorithm calculates the energy levels and their associated degeneracies via a non-Markovian Monte Carlo walk through the system's energy spectrum. The computational time to do this depends upon the number of energy levels and the degrees of freedom of the problem of interest.  However, a great benefit of the framework is that the landscape only needs to be calculated once. The kinetics are subsequently obtained from the SEAQT equation of motion, which is a system of first-order ordinary differential equations, applied to the energy landscape for a specific initial condition. This is a relatively modest problem and can easily be repeated for any number of different initial conditions.  Furthermore, since the SEAQT equation of motion produces a single kinetic curve for each initial condition and expresses properties and descriptors as expected values, there is no need to repeatedly simulate any particular kinetic path over and over again and average the results, as must be done with traditional KMC approaches. Thus, the SEAQT framework is effectively computationally comparable to a couple of Monte Carlo simulations.

\section{Conclusions \label{conclusions_sec:level1}}
The principle of steepest entropy ascent is applied to a simple energy landscape to describe the kinetics of three related physical processes (sintering, precipitate coarsening, and grain growth) under one framework without assuming the system is in local- or near-equilibrium and without {\em ad hoc} assumptions about the rate controlling mechanisms of the processes. The computationally efficient Replica Exchange, Wang-Landau algorithm is used to generate an energy landscape and the degeneracies associated with the energy levels, while a method is proposed for linking microstructural descriptors to state space.

Once an accurate energy landscape and descriptors are constructed, the SEAQT framework is used to find a unique kinetic path via an equation of motion in the form of a system of first-order, ordinary differential equations. With respect to the comparisons of theory with experiment:
\begin{enumerate}
    \item The SEAQT-predicted kinetics qualitatively agree with the available experimental kinetics for ZrO$_2$ sintering, Al$_3$Li precipitate coarsening, and grain growth in nanocrystalline Pd. 
    \item The predicted kinetics can be brought into quantitative agreement by simply adjusting the SEAQT relaxation parameter, $\tau$.
    \item The kinetic path through state space predicted by the SEAQT framework can be linked directly to the microstructure through one or more descriptors averaged over the occupied energy levels. 
    \item The computational burden associated with applying the SEAQT framework to the three physical processes is limited primarily to constructing the energy landscape; solving the SEAQT equation of motion is straightforward and requires limited computational resources. 
\end{enumerate}

\begin{acknowledgments}
The authors thank an anonymous referee for insightful comments and helpful suggestions. We acknowledge Advanced Research Computing at Virginia Tech for providing computational resources and technical support that have contributed to the results reported within this paper. JM acknowledges support from the Department of Education through the Graduate Assistance in Areas of National Need Program (grant number P200A180016).
\end{acknowledgments}

% The \nocite command causes all entries in a bibliography to be printed out
% whether or not they are actually referenced in the text. This is appropriate
% for the sample file to show the different styles of references, but authors
% most likely will not want to use it.
%\nocite{*}

%\bibliographystyle{plainnat}
%\bibliographystyle{abbrvnat}
%\bibliographystyle{unsrtnat}
%\bibliographystyle{apsrev4-2}
\bibliographystyle{ieeetr}
\bibliography{JaredsRefs} % Produces the bibliography via BibTeX.

\end{document}